# An Empirical Study of Architecting for Continuous Delivery and Deployment

Mojtaba Shahin [a], Mansooreh Zahedi [b], Muhammad Ali Babar [a], Liming Zhu [c]

[a] CREST - The Centre for Research on Engineering Software Technologies, The University of Adelaide, Australia
[b] CREST - The Centre for Research on Engineering Software Technologies, IT University of Copenhagen, Denmark
[c] Data61, Commonwealth Scientific and Industrial Research Organisation, Sydney, NSW 2015, Australia
mojtaba.shahin@adelaide.edu.au, mzah@itu.dk, ali.babar@adelaide.edu.au, liming.zhu@data61.csiro.au

**Abstract**

Recently, many software organizations have been adopting Continuous Delivery and Continuous Deployment (CD) practices to develop and deliver quality software more frequently and reliably. Whilst an increasing amount of the literature covers different aspects of CD, little is known about the role of software architecture in CD and how an application should be (re-) architected to enable and support CD. We have conducted a mixed-methods empirical study that collected data through in-depth, semi-structured interviews with 21 industrial practitioners from 19 organizations, and a survey of 91 professional software practitioners. Based on a systematic and rigorous analysis of the gathered qualitative and quantitative data, we present a conceptual framework to support the process of (re-) architecting for CD. We provide evidence-based insights about practicing CD within monolithic systems and characterize the principle of "small and independent deployment units" as an alternative to the monoliths. Our framework supplements the architecting process in a CD context through introducing the quality attributes (e.g., resilience) that require more attention and demonstrating the strategies (e.g., prioritizing operations concerns) to design operations-friendly architectures. We discuss the key insights (e.g., monoliths and CD are not intrinsically oxymoronic) gained from our study and draw implications for research and practice.

## 1. Introduction

Development and Operations (DevOps) is a relatively new software development paradigm that promises many benefits such as solving the disconnect between development and operations teams, faster development and deployment of new changes, and faster failures detection [1-3]. Continuous integration, delivery, and deployment practices are key DevOps practices to fully realize the promises of DevOps without compromising quality [1]. Continuous Integration (CI) is the practice of integrating code changes constantly (i.e., at least daily) into the main branch and entails automated build and testing [4, 5]. As shown in Figure 1[1], CI is the first and core enabler for adopting continuous delivery and deployment practices [2, 6]. Whilst continuous delivery is focused on keeping software releasable all the time, continuous deployment extends it in order to continuously and automatically deploy new changes into production [7].

Precise definitions of continuous delivery and deployment are often missing in both the literature and industrial circles [5, 7, 8]. Continuous deployment is a push-based approach, by which code changes are automatically deployed to a production environment through a pipeline as soon as they are ready, without human intervention [9]. Continuous delivery is a pull-based approach, in which a person (e.g., a manager) is required to decide *which* and *when* production-ready code changes should be released to production [9]. Figure 1 shows the relationship between these practices and how an application is deployed to different environments [10]. Whilst a production environment is where the applications or services are available for end users, a staging environment aims at simulating the production environment as closely as possible. Continuous delivery and deployment share common characteristics and are highly correlated and intertwined [11, 12]. Hence, we refer to Continuous Delivery and Deployment (CD) as CD practices in this paper.

To support CD, organizational processes and practices may need to change and/or be supplemented by innovative tools and approaches. To this end, several efforts have been allocated to study and understand how organizations effectively adopt and implement CD practices. A number of studies emphasize the importance of choosing appropriate tools and technologies to design and augment modern deployment pipelines to improve the software delivery cycle [13-15]. Other research argues that the highly complex and challenging nature of CD practices requires changes in a team's structures, mindset, and skill set to gain the maximum benefits from these practices [16, 17]. The other areas of interest in CD research are focused on improving automation in testing and deployment, and integrating performance and security into the deployment process [18].

---

[1] Note that some of icons used in this figure are taken from **freepik.com** and **thenounproject.com**

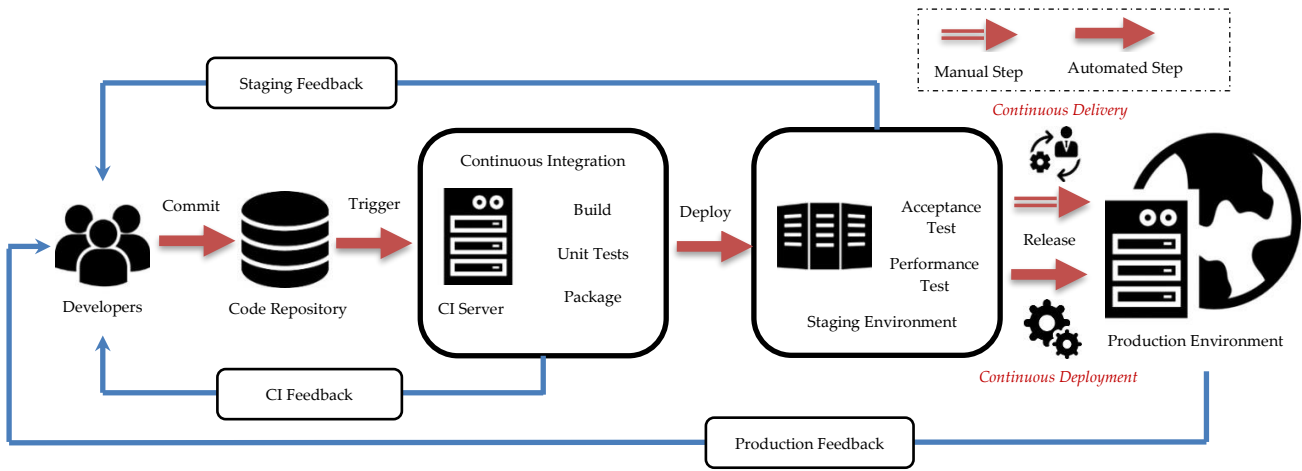

**Figure 1**. The relationship between continuous integration, delivery and deployment

It has recently been claimed that the fundamental limitations to adopting CD practices are deeply ingrained in the architecture of a system and these practices may bring substantial architectural implications [1, 2, 18, 19]. Whilst the industrial community, through white papers and practitioners' blogs, has investigated the role of software architecture in CD adoption [2, 20, 21], there is little empirical effort to study how software architecture is being impacted by or is impacting CD practices [22-24]. This is evident in the recently published systematic reviews on CD [18, 19, 25], in which a new line of research has been called to explore how an application should be (re-) architected for CD. Furthermore, to succeed in the DevOps/CD movement, which emphasizes treating operations teams and operational aspects as first-class entities in the software development process, modern architectures should deal with both design and runtime considerations (e.g., predictive monitoring) [1, 26-29]. For example, software architecture in a CD context should ensure the desired level of quality attributes (e.g., deployability), and reduce the feedback cycle time from operations to development [23]. We assert that an appropriate software architecture (SA) is required to maximize the potential benefits of CD practices. To characterize CD-driven architectures, we carried out a large-scale empirical study, using a mixed-methods approach, which involved 21 semi-structured interviews with participants from 19 organizations and a follow-up survey of 91 software practitioners to assess and quantify the interviews' findings. Our study was guided by the following research question:

*RQ. How should an application be (re-) architected to enable and support continuous delivery and deployment?*

Our results suggest monoliths and CD are not intrinsically oxymoronic. However, adopting CD in this class of systems is more difficult due to the hurdles that they present in terms of *team autonomy*, *fast and quick feedback*, *enabling automation* (e.g., test automation) and *scalable deployment*. To that end, the principle of "*small and independent deployment units*" is used as an alternative to the monoliths by some participants' organizations. We find that *autonomy* in terms of *deployability*, *modifiability*, *testability*, *scalability*, and *isolation of business domain* are the main characteristics of this principle. With that information, we intend to support organizations to move from a monolith to "*small and independent deployment units*". We also discuss quality attributes including deployability, modifiability, testability, loggability, monitorability, and resilience (also indicated by previous research [13]), which require more attention when designing an application in a CD context. We provide concrete examples of these quality attributes in action and discuss their role for CD success. Finally, we demonstrate three strategies (e.g., prioritizing operations concerns) suggested by our participants to design operations-friendly architectures. This study, as the first large-scale qualitative study of architecting for CD, makes the following contributions:

(i) A better understanding of the characteristics of CD-driven architectures;

(ii) A catalog of 16 empirically-justified findings that can be used to help organizations and practitioners to adopt CD and to guide them in creating CD-driven architectures;

(iii) A conceptual framework to support the process of (re-) architecting for CD;

(iv) Concrete and actionable recommendations for both practice and research.

We have previously reported the design and findings of the interview part of our research on this topic, based on an analysis of 16 interviews [30]. This paper significantly extends and enhances the previously published work by adding the analysis and findings from five more interviews, assessing and generalizing the findings from the analysis of the data gathered by surveying 91 practitioners, characterizing "*small and independent deployment units*", and providing additional perspectives on operational aspects drawn from the previously published work.

The rest of the paper is organized as the follows: Section 2 summarizes prior related work. Section 3 presents our research method. We present demographic information in Section 4. The quantitative and qualitative results are described in Section 5. Section 6 closes the paper with our discussion and reflection on the findings.

## 2. Related Work

The related work is divided into the general literature on CI and CD practices, and the specific literature on the role of SA in CI and CD adoption.

### 2.1 General Literature

**Primary Studies:** Transition to CI/CD practices is a nontrivial process that necessitates technical, cultural, process and tooling changes in organizations to support the highly complex and challenging nature of these practices. A set of studies have examined the barriers to implementing CI [6, 31, 32]. For example, Hilton et al. [6] show that debugging CI build failures and long build times are the top two barriers faced by practitioners using CI. Debbiche et al. [31] indicate *test dependencies* and *long-running tests* as barriers in this regard. Another line of research has focused on developing techniques or practices to improve CI adoption (see [18] for more information), e.g., prioritizing tests to reduce build and test times in CI [33, 34] and increasing visibility for CI process through visualizing the data produced in CI [35, 36].

Adams and McIntosh [37] characterize a modern release engineering pipeline by defining six major phases (e.g., infrastructure-as-code), which can help to ease the adoption of modern release practices (e.g., CD practices). The study argues that whilst the industry has started widely implementing these practices, the empirical evaluation of these practices is in the nascent phase. A number of studies have examined the challenges, pitfalls, and changes that organizations may have experienced in CD adoption and/or adopted practices for this purpose. Claps et al. [17] studied the technical and social challenges of adopting continuous deployment in a single case software company. Based on the analysis of the data gathered from 20 interviews with software practitioners in the case company, they identified 20 challenges such as team experience, team coordination, continuous integration, infrastructure, and partner plugins. They reported on the strategies (e.g., rigorous testing for databases) adopted by the case company to alleviate the challenges. The findings of this study reveal that adopting continuous deployment necessitates changing team responsibilities, greater team coordination and involves several risks. Savor et al. [38] present an experience report on implementing continuous deployment at Facebook and OANDA. The study reveals that despite the tremendous increase in the number of team members and the complexity of code size over six years, continuous deployment did not negatively affect the team productivity (i.e., lines of added or modified code deployed to production per developer) and software quality (i.e., number of production failures). Furthermore, the study distills some issues such as management support and extra effort for understanding updates that an organization may encounter during the journey towards continuous deployment adoption. Chen [13] and Leppanen et al. [16] report the benefits that CD practices provide for software development organizations, including reduced deployment risks, lower development and deployment costs, and faster user feedback. A number of obstacles and challenges (e.g., resistance to change, customer preference, and domain constraints) to CD are also reported in [13, 16], which are in line with the challenges reported by Claps [17]. Apart from the challenges reported in [13, 16, 17], the stage-gate process is also a roadblock to continuous delivery success [39]. A stage-gate process has several quality gates to ensure the quality of new releases before entering the next stage. Laukkanen et al. [39] indicate that tight schedules, process overheads and multiple branches, which are associated with a stage-gate process, make it almost impossible to adopt the continuous delivery practice in a stage-gate managed organization.

**Secondary Studies:** Based on a Systematic Literature Review (SLR), Stahl et al. [40] identified 22 variations (e.g., fault responsibility) in the implementation of CI in software industries. From the design perspective, one of the variation points is related to *modularized CI*, discussing the statements about the composition of components in a system, and how components can be managed (e.g., managing source code) independently. Recently, there have been several reviews published on CD practices [18, 19, 25] and rapid release [41]. Mäntylä et al. [41] conducted an SLR aimed at identifying the benefits (e.g., customer satisfaction), enablers (e.g., tools for automatic deployment) and challenges (e.g., time pressure) of rapid release (including CI and CD). Based on 24 primary studies, the review concludes that rapid release is a popular practice in the industry; however, there is a need for empirical studies to prove the claimed advantages and disadvantages of rapid release. Three recently published reviews [18, 19, 25] on CD practices mostly focused on the issues that hinder adopting CD, along with solutions to address those issues. Among others, some of the major stumbling blocks to CD are rooted in the system design, production environment, and testing practices. Furthermore, it has been revealed that the solutions to testing and system design problems in CD are rare. Shahin et al. [18] also define the critical factors that a given organization needs to consider carefully along the CD adoption path, including testing (effort and time), team awareness and transparency, good design principles, customer satisfaction, highly skilled and motivated teams, application domains, and appropriate infrastructure(s). Rodríguez et al. [25] identify 10 factors, such as fast and frequent release, continuous testing and quality assurance, and the configuration of deployment environments, which together characterize CD practices. The configuration management process in CD refers to storing, tracking, querying, and modifying all artifacts relevant to a project (e.g., application) and the relationships between them in a fully automated fashion [42].

## 2.2 Architecting for CI and CD Practices

The software architecting process aims at designing, documenting, evaluating, and evolving software architecture [43]. Recently, Software Architecture (SA) research has been experiencing a paradigm shift from describing SA with quality attributes and the constraints of context (i.e., the *context and requirement* aspect) and structuring it as components, connectors and views (i.e., the *structure* aspect) to focusing on how stakeholders (e.g., the architect) make architectural decisions and reason about the chosen structures and decisions (i.e., the *decision* aspect) [28, 44]. Whilst current research and practice mostly consider architecting as a decision-making process, it has recently been argued that SA in the new digitalization movements (e.g., DevOps) should cover the *realization* aspect as well [28]. The *realization* aspect of software architecture design is expected to deal with operational considerations such as automated deployment, monitoring, and operational concerns. However, the SA research community has provided few guidelines and systematic solutions for this aspect of software architecture [29]. Accordingly, our findings in this study are placed in all four aspects of software architecture in the CD context, as Sections 5.1 and 5.2 mostly focus on the *structure* aspect, Section 5.3 is about the *context and requirement* aspect and Section 5.4 concerns the *realization* aspect. Finally, the decisions and rationales (the *decision* aspect) that are made at the level of architecture for CD, are embedded in these aspects of software architecture.

Nonetheless, whilst previous works indicate that an unsuitable architecture would be a major barrier to CD transition [18, 19, 25, 39], there has been little empirical research on the role of SA as a contributing factor when adopting CD practices. Some initial efforts on this topic have been reported in [22-24, 32, 45]. Mårtensson et al. [32] have conducted a case study to explore the behavior of developers in CI practice in two case organizations. They identify 12 enabling factors (e.g., "work breakdown" and "test before commit") impacting the capability of the developers to deliver changes to the mainline. The study [32] argues that some of these factors (e.g., work breakdown) can be limited by the architecture of a system. Chen [22] reports on the experience of architecting 25 software applications for continuous delivery. The study indicates that continuous delivery creates new challenges for architecting software applications. According to Chen, continuous delivery heavily influences a set of Architecturally Significant Requirements (ASRs), such as deployability, security, modifiability, and monitorability. It is asserted that the maximum benefits of continuous delivery are achieved by effectively satisfying the aforementioned quality attributes. Bellomo et al. [23] studied three projects involving continuous integration and delivery to understand deployability goals of those projects. The study [23] reveals that most of the decisions made to achieve the desired state of deployment (i.e., the deployability quality attribute) were related to the architecture of the systems in those projects. Based on the study's findings, the authors formed a deployability tactics tree. Schermann et al. [24] conducted an empirical study to identify the current practices and principles in the software industry to enable CD practices. They observed that A/B test and dark launches as practices of CD are not often applied in industry and the feature toggles technique may bring unwanted complexity. Recently, a few academic and white papers have discussed microservices architecture as a first and promising architectural style for CD practices [45-47]. Microservices architecture aims to design software applications as a set of independently deployable services [46, 48]. Balalaie et al. [45] report an experience of migrating a monolith to a microservices architecture. Apart from changing team structures (e.g., forming small cross-functional teams), they applied a number of migration patterns (e.g., change code dependency to service call) to decompose the monolithic system into microservices [45]. Furthermore, the migration process included introducing new supporting components and using containerization to support CD practices.

## 3. Research Method

It is argued that the research method should be selected based on the nature and objectives of the studied problem [49]. To help realize this goal, we chose a mixed-methods approach with a *sequential exploratory strategy* [49] consisting of interviews and a survey to find the answer to the research question for this study. Gathering data from different sources (i.e., data triangulation) increases the accuracy and reliability of the results [49, 50]. As shown in Figure 2, after creating an interview guide, we conducted 21 interviews with both architecture and non-architecture practitioners to gather a wide range of perspectives and to gain a deep understanding of how adopting CD practices may impact on the software architecting process. The interviews had a qualitative focus and mainly dealt with "How" and "What" questions (see Appendix A). At the next step, we ran a survey to reach out to a large number of populations of software industrial practitioners. The survey aimed at quantifying, augmenting, and generalizing the findings obtained from the interviews. Although the survey mostly focused on quantitative aspects, we had some open-ended questions to gain further thoughts and opinions from the respondents. We developed a research protocol using appropriate guidelines [51, 52] and meticulously followed it whilst conducting this study.

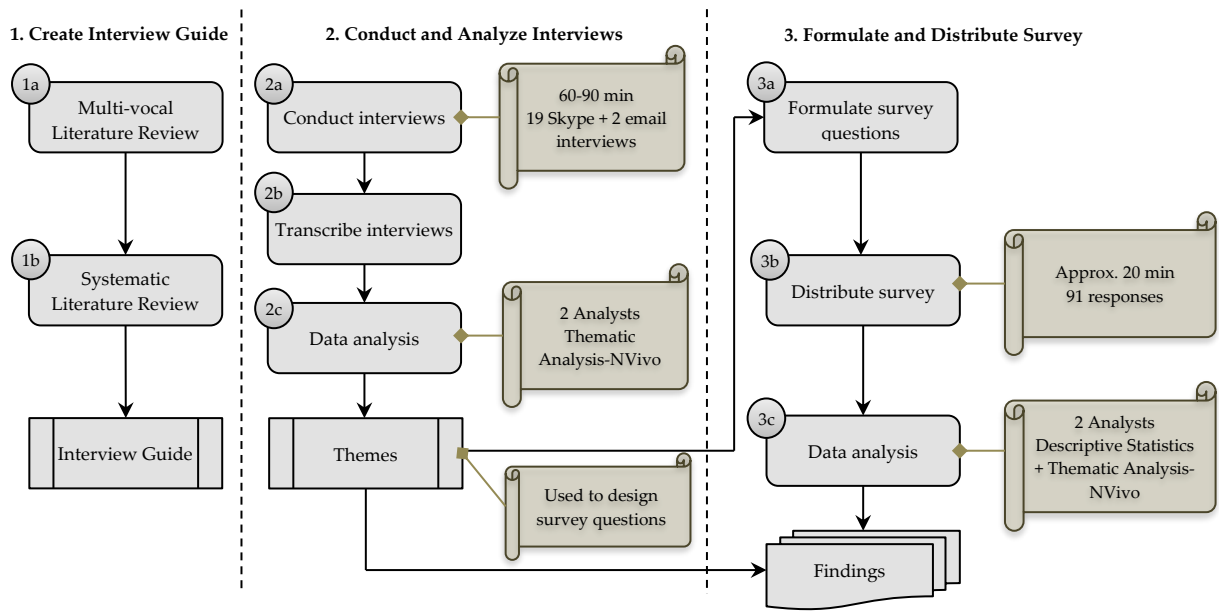

**Figure 2**. Research method steps

## 3.1 Interviews

### 3.1.1 Protocol

The interview guide involved 30 open-ended questions that were designed to support a natural conversation between the interviewer and the interviewee. The interview questions enabled the participants to freely and openly share their experiences and thoughts. We carried out 21 semi-structured, in-depth interviews with participants from 19 organizations. Most of the interviews (i.e., 19) were carried out via Skype. Given the time and geographical restrictions, 2 of the participants opted for sharing their responses via email. After getting practitioners' consent for participation in the interviews, we shared the questions with them before conducting the interviews. This practice enabled the participants to become familiar with the objectives of the interview's questions before participating in the interview [51]. As the study progressed, we modified some of the questions based on the feedback from the interviewees [50]. This included rephrasing three questions, dropping four questions (e.g., the question related to development methodology) to shorten the interview time, and adding two questions (i.e., Q 12.2 and Q 15 in Appendix A). Although the interview questions were only stable after the fifth interview, we believe this did not affect our findings, as the findings reported in this paper were confirmed by multiple interviewees. It is worth mentioning that the interviews were exploratory in nature and generalizability was not the main goal. The first author conducted the interviews, each of which lasted from 60 to 90 minutes. With the interviewees' consent, we audio-recorded the interviews and fully transcribed them for an in-depth analysis. For readability purposes, we corrected the grammatical issues in the transcripts of the interviews. Our interview guide was created based on our earlier systematic review [18] and a multi-vocal literature review (i.e., non-peer-reviewed sources of information such as blogs [53]). The interview instrument had five parts: first, it started with a short explanation of the research goals. Second, the interviewees were asked demographic questions (e.g., role and number of years of experience). Third, we asked them to select at least one project from their organization or client that had adopted or was adopting continuous delivery or deployment practices. In fact, continuous delivery and deployment had to be the major focus in the project to be used as a reference point for each of the interviews. Then the interviewees were asked to share the challenges, pitfalls, and the changes that CD practices may have brought to the architecting process, and the architectural principles and practices that they used to address them. Fifth, we asked them to share how they took into consideration the operations teams and their concerns in their respective development processes. At the end of each interview, we asked the interviewees to share any other comments and potential issues regarding the questions. The interviews produced a total of almost 23 hours of audio and over 90000 words of transcriptions. The key parts of the interview guide have been included in Appendix A of this paper.

### 3.1.2 Participants

We aimed to recruit participants using purposive sampling for this study [54]. We approached software practitioners who either had valuable experience in (re-) architecting for CD (e.g., were the architect) or were closely involved with or influenced by architecture (e.g., were DevOps engineers). They had worked for organizations that had adopted or were adopting CD or were part of CD/DevOps consulting organizations. The interviewees were identified in multiple ways, including industry contacts, our personal networks and (keynote) speakers and attendees of industry-driven conferences on DevOps, CD, or SA. Furthermore, we ran the following search terms on the Google search engine to find highly relevant practitioners in this regard: "*architecting for DevOps*", "*architecting for continuous delivery/deployment*", and "*microservices and continuous delivery/deployment*". Then we invited them for the interview study directly. Apart from rigorously analyzing the potential participants' profiles to understand whether they had the right types of competences for this study, as discussed in Section 3.1.1, the interview questions were sent in advance to the potential participants so that they could decide whether or not they were suitable participants. We motivated the interviewees by giving them a free copy of a book (i.e., "*DevOps: A Software Architect's Perspective*" [1]) after the study. It should be noted that we interviewed practitioners with different levels of seniority, different project roles, and different types of experiences in order to achieve a broad and rich understanding and characterization of the implications of CD practices on the architecting process. The interviewees were working in different types of organizations from a set of diverse domains and of varying business sizes, e.g., from small (<100 employees), medium-sized (100-1000 employees) and large organizations (> 1000 employees). We also used a "snowballing technique" to ask participants to introduce suitable candidates for participation [55]. Out of 21 interviewees, 5 were identified through our personal contacts, 2 using the snowballing technique and 14 by googling and browsing their profiles.

### 3.1.3 Analysis

We performed a qualitative analysis of the interview data using a conceptualized thematic analysis technique in software engineering [56]. Given the large volume of data, we decided to use a qualitative data analysis tool called NVivo[2]. We initiated the analysis by creating three top-level nodes in NVivo: (1) the challenges and pitfalls that the interviewees faced at the architecture level in transition to CD; (2) the architectural principles, solutions, and practices they employed to better support CD; and (3) the strategies and practices that the interviewees' organizations employed to treat operations team and their concerns as first-class entities. Our data analysis process started after the third interview, indicating both data collection and analysis proceeded in parallel [57]. While the first researcher performed the analysis process, the second researcher examined all extracted themes to both confirm the themes and identify any other potential themes. We followed the five steps of the conceptualized thematic analysis as detailed below:

1. *Extracting data*: data analysis began with reading and examining the transcripts of the interviews to extract the *key points* of each interview.
2. *Coding data*: at this step of the analysis, the initial codes were constructed. Our interview mostly targeted "How" questions, which were answered in a few sentences rather than being abstract. This enabled us to extract the initial architectural challenges, principles and solutions as well as strategies for addressing operational aspects (Figure 3.A). Making use of NVivo enabled us to move back and forth between the codes easily and review all the extracted data under a particular code.
3. *Translating codes into themes*: for each interview transcript, we tried to combine the different initial codes generated from the second step into potential themes (Figure 3.B).
4. *Creating a model of higher-order themes*: the extracted themes were constantly compared against each other to understand which themes had to be merged with others or dropped (e.g., through a lack of evidence) [58]. We grouped them into categories that represent a higher-order model of themes (Figure 3.C).
5. *Assessing the trustworthiness of the synthesis*: through this step, we first assessed the trustworthiness of the interpretations from which core themes emerged [56]. In this step, we established arguments for the extracted themes, for example in terms of credibility, are the claimed core themes supported by the evidence of the thematic synthesis? For confirmability purposes, is there any consensus among the researchers on the coded data? Then, each core theme was given a clear and precise name.

Figure 3 shows the application of the conceptualized thematic method on some of the interview transcripts to identify the "*team dependences*" theme.

---

[2]http://www.qsrinternational.com

Preprint - accepted to be published in Empirical Software Engineering (2018)

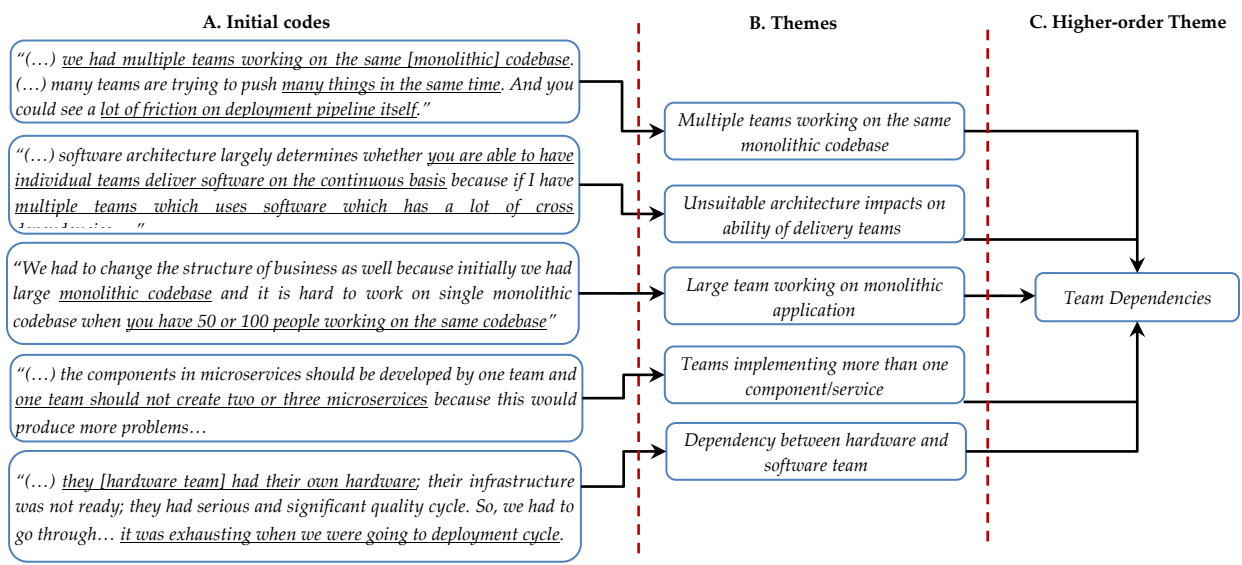

**Figure 3**. The steps of applying conceptualized thematic analysis on the interview transcripts

## 3.2 Survey

### 3.2.1 Protocol

The online survey was designed based on the guidelines suggested by Kitchenham and Pfleeger [52] and hosted on Google Forms. The survey questions were formulated based on the interviews' findings to augment and generalize the findings with a larger sample size. In the survey preamble, we briefly explained the study's goals and the eligibility requirements of potential participants. We also defined the architecting process, continuous delivery, continuous deployment, and deployability terminologies. This information was necessary to ensure that all of the survey participants understood and used those terminologies consistently. Similar to the interview guide, the survey targeted questions about the participants' backgrounds, the impact of CD on the architecting process, and operational aspects of CD. Apart from demographic questions, the survey had 30 questions including five-point Likert-scale (25 questions), multiple-choice (1 question), single-choice (2 questions) and open-ended (2 questions) questions. All questions were mandatory. In multiple- and single-choice questions, we also provided an "Other" field to collect further opinions and thoughts from the participants [59]. Likert-scale questions asked the participants to rate three types of statements: (1) how they agreed or disagreed with the statements (i.e., from *strongly agree* to *strongly disagree*); (2) how important (i.e., from *very important* to *unimportant*) the statements or the challenges reported in the statements were; (3) how frequently the statements occurred (i.e., from *almost always* to *never*). At the end of the survey, we included an optional open-ended question to collect any general comments about the questionnaire. The feedback provided by the participants through this question helped us to rephrase five questions (e.g., removing ambiguity in a question's wording) and add two questions including statements S1 (n=42) and S17 (n=83) in the middle of running the survey to cover more aspects of CD and SA. It is worth noting that the survey questions were not reworded any further after receiving the tenth response. The survey was in English and took about 20 minutes to complete. The complete list of the survey questions is shown in Appendix B of this paper.

### 3.2.2 Participants

We advertised the survey to several groups interested in the topics related to DevOps, CD, and microservices on LinkedIn. Then we emailed 4050 GitHub users and asked them to complete the survey. In the email invitation and survey preamble, we asked the participants to forward the survey to any colleagues eligible to participate. We incentivized participation in the survey by offering five copies of a DevOps book (i.e., "*DevOps: A Software Architect's Perspective*") to five randomly selected respondents, who wished to be considered for the draw. However, fewer than 10% of all responses came from the above-mentioned recruitment approaches. We believe that is mainly because our survey needed an advanced level of knowledge and expertise in both SA and CD. Murphy-Hill et al. [60] also revealed that posting the survey on social networks may not encourage a large number of practitioners to participate. Although we only used the email addresses that were publicly available on GitHub to invite the GitHub users, this approach raised minor issues, e.g., a few of them complained about their email addresses being harvested. Therefore, we approached the highly relevant practitioners by following the process used to recruit the interviewees: finding highly relevant practitioners (e.g., speakers at industry-driven conferences on CD, DevOps, and SA), thoroughly analyzing their profiles, and contacting them directly via email. Overall, we emailed the survey to 487 highly relevant practitioners. In the end, we received 96 responses from all the three recruitment methods. All 96 responses were examined to identify careless responses [61]. We found 5 invalid responses by analyzing outliers, examining inconsistencies in response to two related questions (e.g., S4 and S5), and recognizing the same responses to consecutive questions (e.g., S7 to S10) [61]. It should be noted that we abstained from measuring a response rate for our survey due to having a heterogeneous target population (e.g., practitioners might be members of multiple LinkedIn groups).

### 3.2.3 Analysis

We applied descriptive statistics to analyze the data gathered from the closed-ended questions (e.g., Likert-scale questions) [60]. To analyze the open-ended questions, we followed the conceptualized thematic analysis method described in Section 3.1.3. Similarly, the first author conducted the analysis process and then the second author examined all the extracted themes.

## 3.3 Threats to Validity

Whilst we followed the existing guidelines strictly to conduct this study [52, 62], similar to other empirical studies, there are some threats that may have affected the findings of this study. One of the threats that may occur in any empirical study concerns the sampling method. As we described in Sections 3.1.2 and 3.2.2, we purposively recruited the participants (e.g., analyzing the profiles of potential practitioners). We are confident that most of the interviewees and the survey participants had the right experience and expertise to participate in our study. However, this may have led to an overstatement of the importance of SA in successfully adopting CD [63]. We minimized this threat by recruiting non-architecture practitioners, such as developers and DevOps engineers in both the survey and interviews. We also included participants holding different roles in the same organization to help reduce any personal bias. In the retrospective studies (e.g., interviews), the participants may not have been able to remember all the details during interviews [60]. We adopted two strategies to alleviate memory bias: first we sent the interview questions in advance to the interviewees to help them to refresh their memories of all the relevant details and implicit decisions. We also asked them to share their experiences from their most recent projects or clients. Another threat that may have influenced the participants' answers was social desirability bias [63, 64], in which a participant tries to answer the questions in a manner that s/he perceives a researcher would want. We limited this bias by informing the participants at the beginning of the interviews and survey that personal details are not to be divulged and all the collected data would be anonymized [59]. Another limitation lies in the validity and presentation of the questions used in the interviews and the survey [65]. All the questions were designed by the first author and thoroughly reviewed and validated by the other authors and a select number of industrial practitioners. We also improved the questions' wording based on the received feedback. During the interviews, we mostly used open-ended questions, and extensively encouraged the interviewees to provide as detailed answers as possible. Whilst the email-based interviews may not have generated in-depth information, it is argued that the responses sent for the email-based interviews are more thoughtful and focused [66]. The survey questions originated from the interview findings. Wherever required we included open-ended questions or an "Other" field in order to collect additional information. We note that we did not find too much additional information through the open-ended questions and "Other" fields, suggesting the interviews successfully identified the significant findings.

Researcher bias can be another potential threat to the validity of the findings in a qualitative study. A large part of the data analysis step was conducted by the first author. In order to minimize this threat, the second author investigated all the extracted themes. In case of any doubt, continuous discussions were organized with the second author to maintain the accuracy of the analysis process, which was also guided by the pre-defined research protocol described in Sections 3.1 to 3.2. This study used the triangulation technique to collect data from two sources to minimize any researcher bias. The findings of the interviews and formulation of the survey questions heavily relied on the interviewees' statements, which might be subjective, and can negatively impact on the findings of this study. To alleviate this threat, we have only reported those findings that were confirmed by multiple participants (e.g., at least two participants). In addition, we provided a precise description of the terminologies used in the interview and survey questions to the participants. We are confident that this strategy helped both the researchers and the participants to have a common understanding of the terminologies used.

Similar to other empirical studies, generalizability is a potential threat to the findings of our study. For the interviews, the participants with a wide range of backgrounds (e.g., different roles) were knowingly selected and invited from very diverse types of organizations in terms of size, domain, and way of working, in several countries. We believe our sampling technique largely improved the reliability of our analysis and the generalizability of the findings. Additionally, we augmented and generalized the findings of the interviews through the survey.

## 4. Demographics

### 4.1 Participants' Profiles

We interviewed 21 software practitioners (indicated by **P1** to **P21**) from 19 organizations in 9 countries. Table 1 presents a summary of the demographic details of the interviewees. We also received 91 responses to the survey (indicated by **R1** to **R91**).

**Technical roles**: For both the interviews and survey, the dominant participants were architects (33.3% and 42.8% respectively). The interview study also included 4 consultants, 2 executives (e.g., a Director of Engineering), 2 project managers, 2 technical leads, and there was one developer, one DevOps engineer, one operations engineer and one software engineer. The remainder of the roles participating in the survey included 10 consultants, 9 DevOps engineers, 7 developers, 6 team leads, 6 software engineers, 3 operations engineers, 3 executives, 2 program managers, and 6 with other roles.

**Experience**: 90.4% of the interviewees and 89% of the survey participants had at least six years of experience in software development. Furthermore, 74.7% of the survey respondents had more than ten years of software development experience.

**Organization sizes and domains**: The interviewees and the survey respondents came from organizations of varying sizes. The majority of the participants came from large organizations (i.e., 42.8% of the interviewees, 38.4% of the survey respondents), followed by medium-sized organizations (33.3% of the interviewees and 32.9% of the survey respondents) and small organizations (23.8% of the interviewees and 28.5% of the survey respondents). Although the interviewees and the survey respondents came from a diverse set of organizations in terms of domains, the dominant domains were consulting and IT services, financial services, and telecommunications.

Table 1. Summary of the interviewees' details.

| ID | Role | Exp. Year | Org. Size | Org. Domain | ID | Role | Exp. Year | Org. Size | Org. Domain |
|---|---|---|---|---|---|---|---|---|---|
| **P1** | Architect | >10 | Medium | Research Institute | **P12** | Technical Lead | 8 | Large | Consulting and IT Services |
| **P2** | Developer | 7 | Large | Technology | **P13** | Architect | >10 | Large | Consulting and IT Services |
| **P3** | Senior DevOps Consultant | >10 | Large | Consulting and IT Services | **P14** | Architect | 6 | Medium | Financial |
| **P4** | Program Manager | >10 | Large | Telecommunications | **P15** | Architect | >10 | Small | Consulting and IT Services |
| **P5** | Director of Engineering | >10 | Small | RFID | **P16** | Operations Engineer | >10 | Small | Games |
| **P6** | Vice President of Development | >10 | Large | Financial | **P17** | Technical Lead | 7 | Medium | Games |
| **P7** | Project Manager | >10 | Medium | Networking | **P18** | Consultant | >10 | Small | Consulting and IT Services |
| **P8** | DevOps Engineer | 8 | Large | Telecommunications | **P19** | Solution Architect | >10 | Large | Technology |
| **P9** | IT Consultant | >10 | Medium | Consulting and IT Services | **P20** | Software Engineer | 4 | Large | Technology |
| **P10** | Continuous Delivery Consultant | >10 | Medium | Consulting and IT Services | **P21** | Solution Architect | 2 | Small | Startup |
| **P11** | Architect | >10 | Medium | Consulting and IT Services | | | | | |

## 4.2 Practicing Continuous Delivery and Deployment

We asked both the interviewees and the survey participants to indicate how often, on overage, the applications in their respective or client organizations are in the deployable state (i.e., implementing continuous delivery) and how often they deploy the applications to production (i.e., implementing continuous deployment). These questions were designed, to a large extent, to determine the maturity of implementing continuous delivery and deployment practices (i.e., how an organization adopts and implements continuous delivery and deployment) [16]. It is clear from Figure 4 that 57.1% of the interviewees (12) and 54.9% of the survey participants (50) indicated that on average the applications in their respective or client organizations were in a releasable state *multiple times a day* or *once a day*. This number for continuous deployment practice was lower, as 7 interviewees (33.3%) and 34 survey respondents (37.3%) stated that they automatically deploy their applications *multiple times a day* or *once a day* to production. These results indicate that our findings came from reliable sources as the participants' organizations successfully implemented CD practices. Three interviewees (i.e., P11, P12, and P19) had no idea about how often the application changes were in the deployable state (i.e., shown as N/A in Figure 4). Interestingly, 6 participants indicating the changes were production-ready at least *a few times a month* had actual production deployment *a few times a year*. We argue that the reason for this may stem from factors such as domain constraints and quality concerns [67].

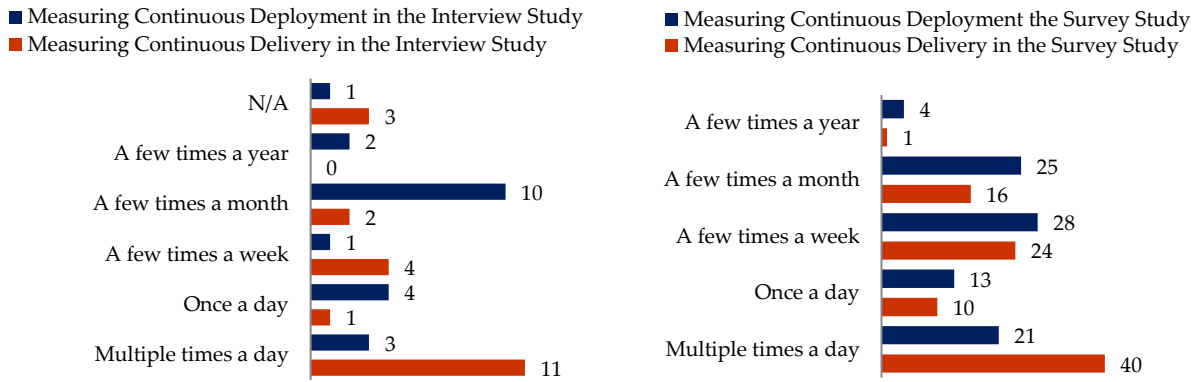

**Figure 4.** How continuous delivery and deployment are implemented in the interviewees' (left) and surveyed participants' (right) organizations

## 5. Findings

Based on the findings from this study, we have drawn a conceptual framework to support (re-) architecting a system for CD (Figure 5). This framework consists of five parts: *Monoliths*, *Migration*, *Small and Independent Deployment Units*, *Quality Attributes*, and *Operational Aspects*. *Monoliths* part is shown on the top left side of Figure 5, which reflects the possibility of practicing CD within monoliths with potential challenges that may hinder CD adoption in this class of systems (Section 5.1). Since monolithic architecture is predominant in software industries [68], there may be organizations that want to achieve CD with their monoliths. To this end, they need to augment/improve the architecture of their systems by applying the practices and strategies presented in *Quality Attributes* and *Operational Aspects*. *Quality Attributes* and *Operational Aspects* have the main goal of creating a CD-driven architecture, apart from the chosen architecture style. It is worth mentioning that this is mainly because the practices and strategies in these two parts were often reported by our participants as prerequisites to, or supportive of, CD-driven architectures. The framework indicates that the *Operational Aspects* can be injected into the architecting process in the context of CD and *Quality Attributes* serve as input to both *Monoliths* and *Small and Independent Deployment Units*. The *Operational Aspects* (left bottom side of Figure 5) provide the strategies to design operations-friendly architectures (Section 5.4). At the bottom of Figure 5, we have the *Quality Attributes* that need to be carefully considered to design CD-driven architectures (Section 5.3). We find that these quality attributes impact the architecture in the CD context in two dimensions: positive (+) and negative (-).

The challenging nature of the monoliths may compel organizations to move from monolithic systems with long development and deployment cycles to *Small and Independent Deployment Units* (top right side of Figure 5). The framework supports organizations in this journey by providing a list of reliable factors, shown as *Migration* in the middle of Figure 5. They can be used to characterize *Small and Independent Deployment Units* (Section 5.2.1). The migration journey results in vertical layering or microservices (Section 5.2.2).

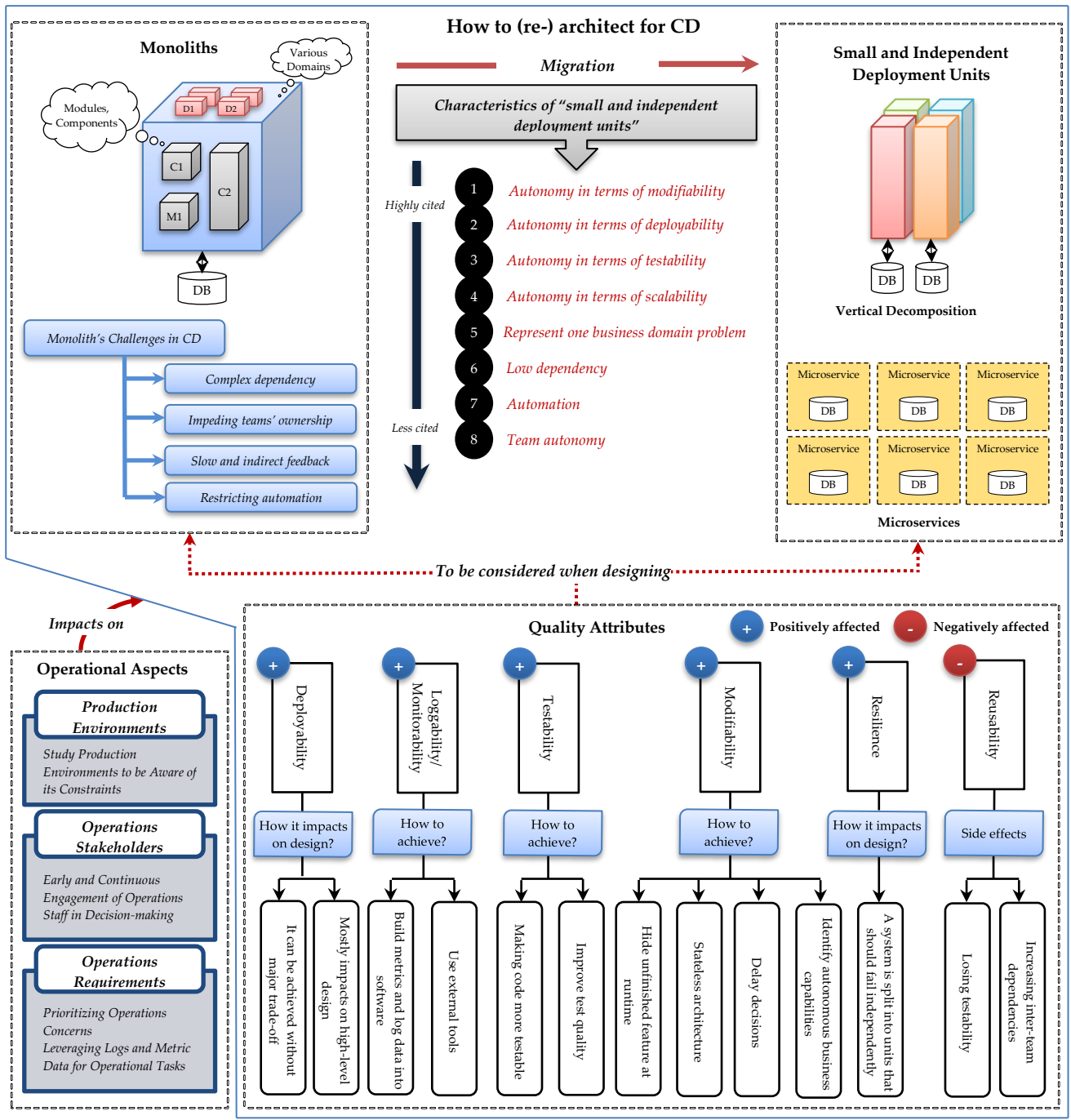

**Figure 5**. A conceptual framework of the findings showing how to (re-) architect for CD

## 5.1 Monoliths and CD

In its non-technical definition, a monolith is defined as "*a large block of stone*". Exploring literature and practitioners' blogs shows that there is no common understanding and interpretation of the term monolith as different types of monoliths may be created during software development: for example, monolithic applications, monolithic builds and monolithic releases [69]. In our study, we define a monolith as a single executable artifact that contains various domains, layers, and many components, modules, and libraries, in which all the functionality is managed and packaged in one deployable unit (see Figure 5) [70, 71]. Our interview study demonstrates that monolithic architectures are seen as a major problem in organizations during CD adoption. The following is a representative quotation about the negative impressions given of monoliths: "*I do not know [if] they [the architects] are aware of that; they create kind of monolithic applications and the monolithic application contains large functional domains [that] would be hard to use in a large-scale organization if you want to have continuous delivery*" **P10**.

Whilst the monolithic architecture, as a barrier to CD, was one of the most commonly occurring codes in the interviews (i.e., indicated by 15 interviewees, of which 6 of them were in the role of architect), we also found evidence in the interviews that CD can be implemented in monoliths. Particularly, when there is a loosely coupled and modular architecture with clearly defined interfaces and one single team working on it, e.g., *"I have seen [examples of adopting] CD in monolithic applications, when they [organizations] don't split them and they go very fast because [there is] still single team, they can be extremely fast, high-quality. The software is modular, and it is not split, and it works well for them"* **P14**.

Furthermore, while conducting the survey, we received comments from the survey respondents that they had been successful in adopting CD within monolithic applications (e.g., *"A monolithic service can use CD"* **R44** and *"Microservices are not required nor is breaking up a monolith [for CD]"* **R22**). All this motivated us to add a new statement to the survey: *"It is possible to practice CD successfully in monolithic applications"*. Figure 6 shows that 59.5% (25 out of 42) of the survey respondents answered this question as *agree* or *strongly agree*. Only 21.4% responded *strongly disagree* or *disagree* and others (i.e., 19%) took a *neutral* position in this regard (see statement S1 in Figure 6). It is interesting to note that among 25 respondents who believed in the possibility of implementing CD within monoliths, only 8 were in the role of architect. 6 architects (strongly) disagreed with statement S1, and 2 adopted a neutral position. Team leads, consultants, and developers were the roles that were the most positive about monoliths. We assert that this finding is not necessarily in conflict with the interviews' findings, as we did not conclude that it is *impossible* to practice CD within monoliths, but it seems that it would be much harder and more complicated to achieve CD within monolithic applications (e.g., *"Any component or service can adopt CD: a larger one [has] a slower cycle. Components without tests cannot adopt CD"* **R85**).

In line with the work by Schermann et al. [24], our interview study shows that once the size of an application grows (e.g., by expanding its functionality) and the number of teams increases, the monoliths significantly impeded achieving scalable continuous deployment (e.g., *"The main challenge is the weight of your architecture, it could be the reason why you can't move to CD"* **P15** and *"We had to change the structure of our business as we had a large monolithic codebase and it was hard to work on when you have 50 or 100 people working on the same codebase"* **P14**). This is mainly because the monoliths may slow down the deployment process and a small change may necessitate rebuilding, retesting and redeploying the entire application [72, 73]. The next subsection lists the main categories of challenges about monoliths, which together hinder achieving CD.

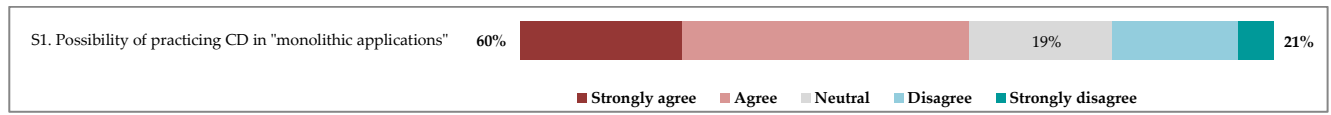

**Figure 6**. Survey responses to the statement on the possibility of practicing CD within the monoliths (n=42)

### 5.1.1 Why it is difficult to practice CD within the monoliths

**C1. Dependency (Tightly Coupled):** As the size and complexity of a monolith grows, the dependencies across the system will be so strong that it will become difficult to change different parts of the monolithic system independently. This impedes deploying software on a continuous basis, as there is a need to thoroughly analyze and maintain all the dependencies in the deployment process [74]. One interviewee described this point vividly, *"When you say we have software component X and I would like to deploy that to the customer; do I need to deploy other software components as well; do I really need to or not? These are the design issues that we really face"* **P9**.

External dependencies with other applications were another roadblock to frequent deployment (i.e., also confirmed by 24.2% of the survey respondents). Whilst an application might always be at a releasable state, deploying a modified functionality within that application on a continuous basis may also necessitate the deployment of all dependencies (e.g., dependent applications). Therefore, software organizations need to refactor other legacy applications and ensure that there is no integration problem in the deployment process. As explained by an architect, *"You always get integration challenges [in the deployment process] because you're mixing new code, new systems with old systems; you still rely on old systems to do something"* **P13**.

Our study identifies challenges regarding (monolithic) databases in the context of CD. The interviewees' organizations often faced difficulties in deploying monolithic databases continuously as modifying any functionality in an application demands changing and incorporating database schema as well. Therefore, they gradually become an operational bottleneck and an undeployable unit. The interviewees expressed dissatisfaction with monolithic databases, e.g., *"One of the traditional [approaches] is that for big applications, they [organization] used one database. Then you know for every piece of functionality that you change, you also need to incorporate database changes, you need to test the database. These are a source of single point of [failure] for these big organizations, so they cannot continuously deploy the database. So, they need to refactor [it] and do changes [in] the software architecture"* **P9**.

Our combined findings indicate that software organizations often need to split their existing monoliths into small deployable parts that can be maintained and deployed independently for supporting CD. However, the interviewees highlighted the difficulties of breaking down a monolith (at application and database level) into smaller units. Through the survey, we asked the participants the extent to which they understand this issue as a challenge. As can be seen in Figure 7, 82.4% of the respondents considered the difficulty of splitting a (monolithic) application into independently deployable and autonomous components/services (i.e., statement S2) as *very important*, *important* or *moderately important* challenge. Furthermore, the difficulty of splitting a single-monolithic database (i.e., statement S3) was also widely recognized by the participants, as only 23% of the survey respondents considered it as *unimportant* or *of little importance*.

**C2. Teams**: According to Conway's law, the architecture of a system mirrors the communication structure of the organization that designed it [75]. Our analysis demonstrates that introducing CD is challenging in those organizations where multiple teams work on a monolithic application. We observed that <u>cross-team dependencies</u> could cause frictions in the CD pipeline. While one team could have full control over its development process, run quality-driven builds and release its output constantly, they can still be dependent on the performance of other teams. Therefore, they can easily break the changes that other teams are working on. As **P12** explained, "*We had multiple teams working on the same [monolithic] codebase. We started noticing that it was really affecting our ability to deploy software [continuously], because many teams were trying to push many things at the same time*". The interviewees also referred to tension between software and hardware teams as a challenge to truly practicing CD, as software teams sometimes rely on infrastructure readiness at production, "*Integration issues were part of the deployment [process] and [we had to] deal with their [hardware team's] backend systems. They had their [own] infrastructure, [which] was not ready; they had a serious quality cycle. So, we had to go through this [cycle], which caused additional preparation work. It was exhausting [for us] when we [as the development team] were going to the deployment cycle. So, it was a challenge to adopt DevOps*" **P5**. Furthermore, the interviewees' organizations that had large teams working on the same codebase experienced difficulties in <u>coordinating</u> among interdependent teams and <u>planning</u> before each release.

As shown in statement S4 in Figure 7, 70.3% of the survey respondents ranked the challenge of <u>team interdependencies</u> and <u>coordination effort</u> when adopting CD as *very important* or *important*. CD may also require changes in organizational structures in order to align them with CD throughput [76]. When we asked the respondents how important the challenge of "*inflexibility of the organization's structure with the spirit of CD practice*" is, 69.2% of them ranked this as *very important* or *important* (see statement S5 in Figure 7)

**C3. Feedback**: Difficulties in getting fast, direct, and optimum feedback on code and test results in the monolithic applications was another category of challenges. Our analysis revealed that <u>long build time</u>, <u>long-running tests</u>, and <u>size of changes</u> in the monoliths were the main causes of slow and indirect feedback. As **P12** explained, "*[In our monolithic architecture] we started noticing that the number of tests started loading up, the feedback cycle was becoming very slow. We arrived at a state where [we had] very little ownership of [things] like test failures. We would have taken, like, around 3 or 4 hours to get any feedback. You could see a lot of friction on the deployment pipeline itself*".

**C4. Automation**: A key practice in CD is automation [42]. A few interviewees 'experience suggested that the heavy-weight nature of monoliths can often be a challenge (e.g., extra effort is needed) to fully automate tests and deployment across different environments. For example, **P15** explained how this situation would result in a longer and slower CD pipeline and extreme difficulties to move to automation: "*If your monolithic architecture is complex, it may be hard to move to continuous delivery because the more components you have, the harder it is to install them, to deploy them, the pipeline will be longer, and harder to automate*".

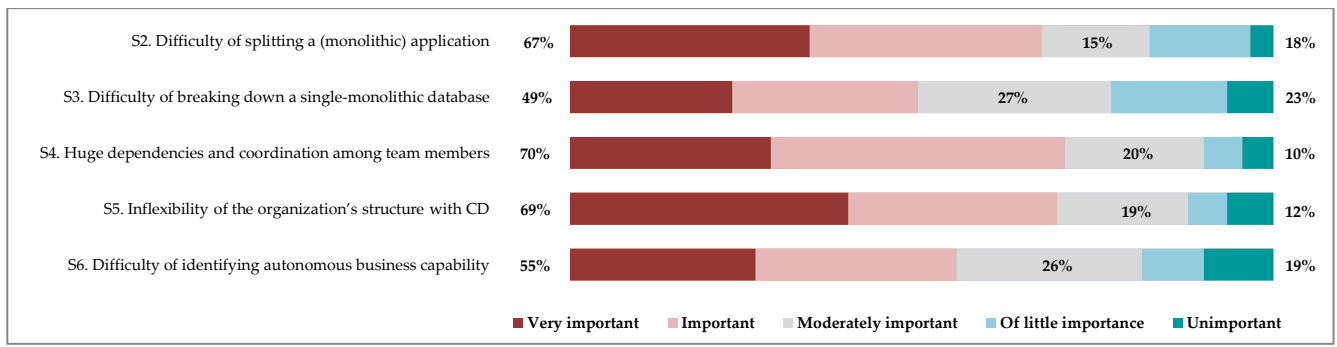

**Figure 7**. The survey respondents indicated the most important challenges in architecting for CD. Inflexibility of the organizational structure with CD is the most critical barrier for implementing CD.

> **Key Findings**
>
> *Finding 1.* Monoliths (application and database) are great sources of pain for CD success, as they are hurdles for having **team autonomy**, **fast** and **quick feedback**, enabling **automation** (e.g., test automation) and **scalable deployment**. Furthermore, among other roles (e.g., consultants), software architects are more pessimistic about CD success within monoliths.
>
> *Finding 2.* Implementing CD is a challenge where multiple teams are working on one monolithic application, as they gradually become **dependent** and need to spend too much time to **coordinate** and **plan** the delivery process.
>
> *Finding 3.* The inflexibility of the organizational structure with CD is the most critical barrier for implementing CD.

## 5.2 Moving Beyond the Monoliths

In the previous section, we discussed how practicing CD in large monolithic systems with multiple teams is not a straightforward approach, as it might have negative impacts on team autonomy, direct and quick feedback, and automation. In this section, we first characterize a key architectural principle adopted by the participants to address the reported monolith-related challenges and then investigate how the key principle is implemented in industry.

### 5.2.1 Characterization of an Architectural Principle: *Small and Independent Deployment Units*

Our findings agree with the argument of Lewis and Fowler [48], that a key architectural principle employed by the participants' organizations for successfully practicing CD is to design software-intensive systems based upon "*small and independent deployment units*" (e.g., service, component, and database). By 'architectural principle', we mean fundamental rules and approaches that serve as a guide to architects to govern architecture designs [77]. **P18** described that applying this principle enabled them to bring the small units into production independently, "*We are moving to smaller services, where microservices to my mind are a less interesting aspect, but the services are much smaller than in the past and they will be deployed into their own server and they [are] managed [e.g., deployed] by one team*".

As we described in Section 5.1, the participants' organizations had challenges with splitting the monoliths into smaller chunks. That is why they usually perform this process incrementally. As an example, "*We started with a big monolith. To be honest, initially we simply split it into a number of smaller chunks; maybe like seven or eight. That already gives us more value [in the deployment process]. In the future, we might split things further*" **P18**.

Gradually breaking down applications should avoid the difficulties of service granularity. **P13** shared that it is easier to initiate the decomposing process with a few large services and incrementally decompose them into smaller, fine-grained units: "*If you are trying to go from monolithic to something like service-based or microservices, don't try to break it down into little tiny pieces first of all right away. You'll get struggles at service level, structurally you have to look at how coupled your components are*".

Nevertheless, the above principle leaves us with yet another question: what does it actually mean? To characterize this principle, we asked the interviewees a number of questions to elaborate on their perception from a small and independent deployment unit, which serves as the foundation for CD success. We found that the practitioners typically consider four main criteria to characterize small and independent units, and accordingly design applications or break down large components or monolithic applications for practicing CD. These criteria include autonomy in terms of **deployability** (e.g., "*[The monolith is split into] those small components [that] can be deployed and reproduced more quickly so as to map smaller iterative deliverables*" *P5*), **scalability** (e.g., "*The majority [of criteria in the decomposing process] were [to have] components that are very scalable, which means whatever you design and whatever you develop must be scalable*" *P5*), **modifiability** (e.g., "*[The monolith is] split into components, so that changes are likely to influence just one component*" *P11*), and **testability** (e.g., "*It can be tested/qualified by itself, and won't break a product or require other services/components to be pushed*" *P19*). Furthermore, some other criteria emerged from data that were only reported by a few interviewees. To give an example, for one of the interviewees *small* meant that the maximum amount of work required for coding and testing a single feature should not exceed a three-day effort for one person.

We asked the survey respondents to rate these four top criteria (see statements S7-S10 in Figure 8) with a 5-level Likert-scale. The results show that, in general, the respondents considered all the above-mentioned criteria for this purpose. As shown in Figure 8, for each of the statements, fewer than 10 respondents disagreed or strongly disagreed. However, statement S7 ("a component/service is small if it can be modified (changed) independently") received more attention than others, and 78%, 76.9% and 68.1% of the survey respondents answered to statements S9, S8 and S10 *strongly agree* or *agree,* respectively.

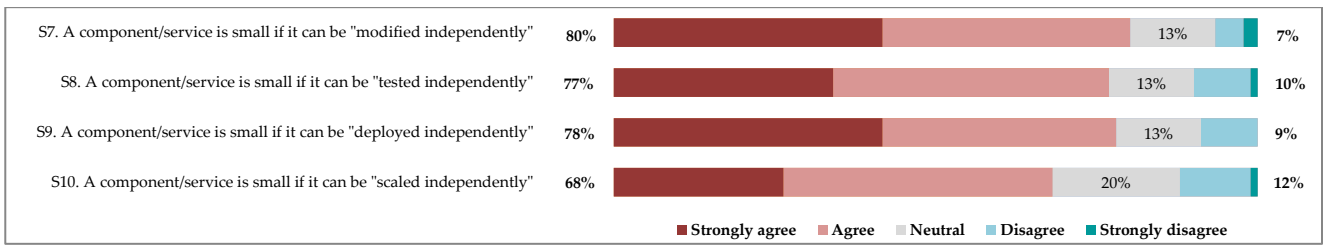

**Figure 8**. The main characteristics of the principle of "small and independent deployment units"; A few survey participants (less than 12%) disagree with the above-mentioned criteria.

Through a mandatory open-ended question, we also asked the surveyed practitioners to share any other criteria or factors (i.e., those not covered by statements S7-S10) that need to be considered to characterize small and independent units for CD and also to decompose the monoliths for this purpose. Some of the respondents tried to elaborate what they had previously chosen in the Likert-scale questions (e.g., "*It [the service] can be deployed/updated independently of any other service*" ***R84***, "*Every feature (roughly mapping onto an Epic) should be small enough to be deployed independently*" ***R31***). The analysis of the open-ended question revealed four other criteria or factors as listed below, in order of their popularity (see Figure 9):

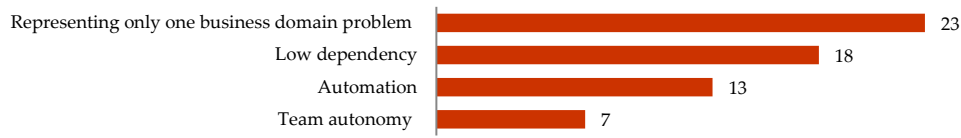

**Figure 9**. The additional criteria/factors shared by the survey participants to characterize "small and independent deployment units"

**Representing *only* one business domain problem (23):** Although a few interviewees indicated that monoliths need to be decomposed into smaller units, so they only cover and solve *one business domain problem*, this criterion was highly cited by the survey participants. For example, **R88** explained, "*A component [in a CD context] should implement a frontend and backend for one business concept*". We found that the definition of *one domain problem* varied significantly among the respondents. For some respondents, it means one function or one task (e.g., "*[A small deployment unit in CD] should encapsulate one unit task end-to-end*" ***R20***), however, others believed that it can serve a logical set of functionalities for a line of business capability, e.g., "*One [service] that solves one specific domain problem, not too small to be just a function, not bigger to try to solve different domain's business*" ***R43*** and "*It [the service] should manage one aspect or collection of features [within] a bounded context*" ***R41***.

The participants suggested different techniques such as *domain-driven design*, *bounded context*[3], and *event storming* to identify a business domain problem. For example, one of the interviewees told us: *"We primarily use techniques such as domain-driven design and event storming to actually identify the autonomous business capability of software architecture"* ***P21.*** Domain-Driven Design (DDD) aims to "design software based on connecting the implementation to an evolving model of the core business concepts" [78]. Bounded context is a concept in DDD to describe the conditions under which a specific model is defined [78, 79]. The models in each bounded context do not need to communicate with the models inside other contexts [46]. Event Storming is a collaborative design technique in which all the key stakeholders (e.g., domain experts) assemble to identify and describe what operations (i.e., domain events) happen within a business domain [80].

**Low dependency (18)**: Some participants stated that the size of a component/service is not an important factor for them, but a component or service should have little dependency on other components/services, in particular during the deployment process: "*[It is] less about size (e.g., lines of code), [but I think] more about loose coupling and clear implementation for component APIs, such as RESTful interfaces*" ***R46***. Respondent **R91** emphasized that a service in a CD context needs to be decoupled from other services in such a way that the upgrade and downgrade of the service would have no impact on SLAs and ongoing transactions. This factor can be augmented by implementing a well-defined interface for each service or component (e.g., "*[A component should be a] small code base, independent of all other services except via API calls*" ***R72***). A few participants also reported that having separate data storage per service (i.e., no common database) can help to achieve low dependency among services. Respondent **R61** stated that a service should have "*no shared storage (with other services)*", while another respondent pointed out that "*my definition isn't about the size, it's [about] the boundaries and responsibilities, including what data it owns versus what data it uses*" ***R39***.

**Automation (13):** Another factor that emerged from the responses was that a small and independent unit needs to be tested and released in an entirely automated fashion within a time-boxed slot, in which it does not require to coordinate with external components/services: "*It must be designed with the intention of being easily built and automatically tested and deployed*" ***R53***. The respondents shared that CD essentially requires components or services with fast build times and fast testing loops in order to have quick and direct feedback, e.g., "*The service should be built and tested in less than a couple of minutes*" **R76**.

---
[3]http://martinfowler.com/bliki/BoundedContext.html

**Team autonomy (7)**: We perceived that having team-scale autonomy strongly affects the size of a component and decisions during decomposition. The respondents shared that adopting CD practices for a component/service is easier if *one* team can comprehend, build, test and deploy it (e.g., "*[A component should be] small enough for the owning team to comprehend it*" **R9**). Another participant answered as "*A team should be able to own a backlog of multiple deployable [items]*" **R30**.

> **Key Findings**
>
> *Finding 4.* The participants' organizations are increasingly considering **small and independent deployment units** as a key architectural principle to provide them with more flexibility in a CD path and achieve frequent and reliable deployments.
>
> *Finding 5.* Autonomy in terms of **deployability**, **modifiability**, **testability**, **scalability**, and isolation of the **business domain** are the main factors to characterize the principle of "small and independent deployment units", and they significantly drive decomposing strategies to **safely** and **incrementally** break down a monolith into smaller independent parts.

### 5.2.2 How is the principle of "small and independent deployment units" implemented in industry?

To embrace the principle of "small and independent deployment units", the participants' organizations usually adopted two approaches: vertical layering and microservices (see Figure 5). Whilst these approaches might be used in many different scenarios to reach different goals (e.g., scalability), we examine them from the perspective of CD practices (i.e., promoting delivery speed). We note that vertical layering and microservices share many common characteristics. As shown in Figure 5, both represent independent deployment units, but the main difference between them is the level of granularity. Vertical layering is more coarse-grained than microservices [70, 81]. Verticals are autonomous systems that are larger than microservices. Verticals are mainly used to reduce operations' complexity, as there would be fewer deployment units. Another notable difference is that vertical layering allows sharing of some assets such as databases and infrastructures, as exemplified by **P17**, "*Every vertical slice would have talked to the same databases in the same way unless we explicitly decided that we didn't want one service talking to a specific backend. It was a hybrid model that worked very well for us*".

**Vertical layering (decomposition)** was a significant approach attempting to embrace the principle of "*small and independent deployment units*". Through this approach, a software application or a large component is decomposed into vertical layers (i.e., independent, autonomous systems) rather than horizontal layers; accordingly one team (ideally) would be responsible for one layer or component during the full development lifecycle [70] (e.g., "*We are changing the paradigm of a more horizontal layered architecture to a more vertical one and isolating architectures, as it will help us to get this performance [deployability]*" **P9**). Vertical layering also decreases interdependency among teams as a team minimally depends on what other teams are working on or are responsible for. Furthermore, this approach hampers the ripple effects resulting from any changes. Nevertheless, it was found that adopting this approach was associated with challenges in the formation of one integrated development team comprising all the required skills (e.g., operations skills), as there is a tendency in software organizations to team up those human resources with similar skill sets. Here are just a few of the examples indicating the benefits of this approach for CD:

"*We had separated layers in the architecture. If there are three teams responsible for one of these layers, all teams need to work together to bring a new piece of functionality alive [in production]. To make it better [the delivery process], we swapped the organization and now each team is focused on one functional domain and works in its own layer. So, they are working in large isolation and we minimize the dependencies*" **P10**.

"*We also changed the teams to align them to one or two (autonomous) subsystems. It was a big change because before that the teams worked on any part of the system. So that contributed to the monolith. It's not perfect but we definitely get some benefits from splitting down the domain lines and restricting development of that part to just one team as they can focus on it and have ownership around that*" **P18**.

**A microservices architecture style** was adopted by some interviewees' organizations as a practical architectural style that suits CD [1, 82]. Our interviews revealed that some of the organizations have been able to successfully adopt the microservices style to smooth their CD adoption journey. Nevertheless, it was evident that implementing this architectural style was also associated with challenges that could negatively impact on the deployment capability of organizations. As mentioned by **P15**, "*So microservices solve some of the issues but also introduce some other issues, especially the orchestration and the configuration of that. Microservices can bring some overhead for operations team as well… if you split too much, you have too many microservices. The operational aspects [of] managing all those services becomes more complex because you have to make all the configuration options available, you have to orchestrate all those services*".

Due to these difficulties, some of the participants' organizations either avoided implementing microservices or did not experience promising results from this architectural style. One of the frequently raised issues regarding microservices was the operational overhead that they introduce (e.g., for monitoring and administrating services), which require highly skilled operations teams for implementation. **P12**, a technical lead and architect, opted to break a monolith codebase into smaller components, each with its own repository. Whilst this decision gives more flexibility to the teams working on the codebase, as the teams were not mature enough to handle multiple runtime services, they decided to have one deployment unit rather than several runtime services. As stated by **P12**, "*We felt that the teams were not ready for that [implementing microservices] and then we compromised, deciding we are not going to make the application as microservices at least for now*". Each component, which in this case is a *binary dependency*, is added as a *runtime dependency* on the parent application. We recognize that the implementation of a microservices style of architecture enforces changes in the organizational structure. The organizations that successfully implemented the microservices style tended to have the structure of teams and their communication pattern aligned with this architectural style. Yet, being able to implement the microservices style at this level requires team maturity and organizational readiness [83]. **P14** observed, "*A company struggles with microservices architecture and they have implemented microservices, as they slow down [software releases]. [It is] because they are not ready to change the organization to support this microservices architecture. The microservices' interactions don't reflect the structure that they have*".

We have previously, in Section 5.1, discussed how (monolithic) databases bring unique challenges to CD. Two of the above-mentioned approaches highlight the importance of revising the core database design and decomposing it into <u>smaller individual databases</u>. There are repeated statements in the interviews which support the need to treat the database as a <u>continuously deployable unit</u>, similar to other software components. <u>Incorporating the database in the CD pipeline</u> as a software component is expected to avoid unexpected issues that database updates may pose at production deployment. We found that <u>keeping everything related to databases (e.g., associated configurations) in a version control system</u> in a consistent manner plays a crucial role in improving the deployability of databases, because that helps trace the changes to the database schema. **P18** emphasized that this practice "*is really useful to monitor databases in different environments and [to] compare them to know [whether] databases have deviated [from] version control*". Using <u>tools to automate</u> the database schema changes and configurations, and also to automate detection of database changes in different environments, was also indicated as helpful for continuously deploying databases. Furthermore, our findings show that a <u>schema-less database</u> (i.e., the schema exists in code, not in the database) is more compatible with the spirit of CD practices. As mentioned by participant **P17**, "*Schema just exists in the code instead of living permanently in the database. Schema-less lets you change the schema with a code, which is exactly what you want with continuous deployment. This is a significant architecture change driven almost exclusively by continuous deployment requirements*".

---

**Key Findings**

*Finding 6. Vertical decomposition and microservices are two primary architectural styles to implement the principle of "small and independent deployment units" in industry. However, adopting these architecture styles comes at a cost as they necessitate considering organizational structures and highly skilled teams (e.g., operations skills). Ignoring this necessity may negatively impact the deployment capability of an organization.*

*Finding 7. The key practices to improve the deployability of the database in a CD context are (1) incorporating the database in a CD pipeline as a software component; (2) keeping database configurations in version control; (3) automating database schema changes and migration, and (4) using the schema-less database.*

---

## 5.3 Quality Attributes that Matter (*Really/Less*) in CD

While we have investigated both practicing CD within monoliths and breaking apart the monoliths into "*small and independently deployment units*", we are also interested in the quality attributes that are most likely to affect (negatively or positively) CD success (see Figure 5). In the following sections, we try to answer this question regardless of the choice of architectural styles, as the following quality attributes deserve serious consideration to realize the anticipated benefits from CD.

### 5.3.1 Deployability

The findings from our interviews indicated that deployability has gained a high priority in CD as the interviewees frequently shared that deployability concerns are accredited during architectural design and drive many decisions to have independently deployable units [2, 22]. We were interested in understanding how deployability concerns impact different aspects of (architecture) design. Taking inspiration from Manotas et al. [84], we introduced statements S11-S14 to ask the surveyed practitioners how often deployability concerns impact the *design of* individual classes, components/services, interactions among components/services and the entire application. We provided a definition of deployability for all the participants to ensure common understanding. We defined deployability as: "*deployability is a quality attribute (non-functional requirement) which means how reliably and easily an application/component/service can be deployed to a (heterogeneous) production environment*". Figure 10 indicates that the frequency of impacting deployability concerns on low-level designs is not significant as a sustainable number of the respondents (70 out of 91, 76.9%) indicated that in the projects adopting CD practices, deployability impacted class's design *sometimes*, *rarely* or *never*. In contrast, high-level designs were more influenced by deployability concerns (see statements

S12-S14 in Figure 10). It is a commonly held belief among over 60% of the respondents that deployability impacted the design of components/services, interactions, and entire applications *often* or *always*. We observed that none of the respondents answered statements S12 and S13 as *never* and only 3 respondents believed that the design of the entire application was *never* impacted by deployability concerns.

During the interviews, we found that deployability has a minimum conflict with other quality attributes (e.g., "*They [quality attributes] sometimes have conflicts; deployability has a minimum conflict with other quality attributes*" **P4**). Another interviewee, **P14**, described the relationship between deployability and other quality attributes as follows: "*I would say that deployability doesn't really conflict with other aspects of continuous delivery, but it requires some practices*". To investigate this claim further, we asked the participants how frequently they could compromise other quality attributes to improve the deployability of an application. Figure 10 shows that the vast majority of the respondents are not willing to compromise other quality attributes to improve deployability. 68.1% of the respondents answered statement S15 with *rarely* or *often*, while 29.6% believed that they could *sometimes* sacrifice other quality attributes to achieve more deployability.

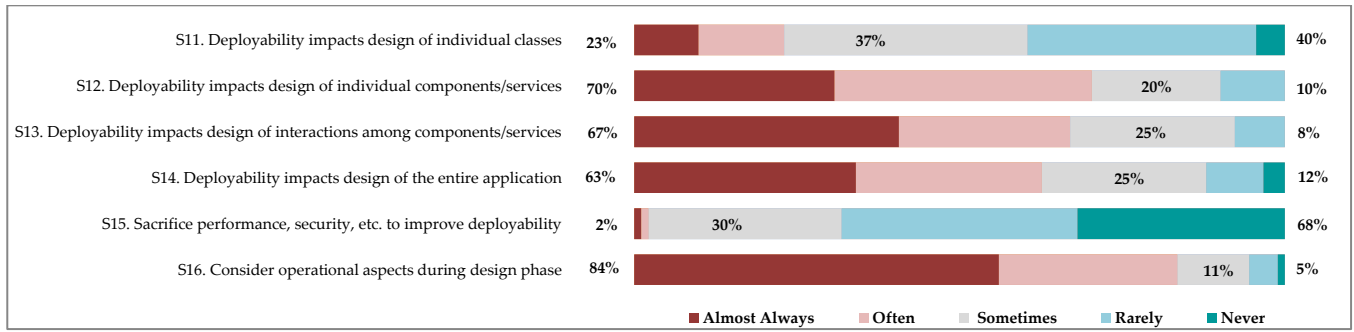

**Figure 10**. Survey responses to the statements on deployability and operational aspects

> **Key Findings**
>
> *Finding 8. Concerns about deployability (e.g., ease of deployment) impact how applications are designed. However, high-level designs, in particular **interactions among components/services**, are more influenced by deployability concerns than low-level designs (e.g., the design of individual classes). These concerns can be supported in CD without major trade-off, as they have minimum conflict with other quality attributes and most of the quality attributes are in support of deployability.*

### 5.3.2 Modifiability

One of the top priorities for the participants when designing an application in a CD context was to support frequent and incremental changes. They reflected that they break down software into units that are small enough to be modified, replaced and run independently. The participants also tried to minimize the impact of changes for CD, as described by one of the interviewees (i.e., an architect) as: "*For me, autonomy is the most important quality attribute for continuous delivery. If you don't have dependency, you can isolate your changes and, as a team, you can independently do your lifecycle and of course, your lifecycle also means bringing it to production*" **P19**. We learned that the participants employ four main techniques for this purpose.

**T1. Identify Autonomous Business Capabilities**: Our findings show there is a tendency among the participants' organizations to structure an architecture based on <u>business capabilities rather than functionalities</u> using techniques such as domain-driven design, bounded context, and event storming (as described in Section 5.2.1) [48]. The main idea behind identifying the autonomous business capabilities of software architecture is to have units that can be independently developed, modified and deployed [85]. However, our interviewees also suggested this would <u>decrease team interdependencies</u>, as mentioned by **P14**, "*[In our CD journey] we needed to break down requirements into different, isolated, and autonomous business values and then each team had to be assigned to them, so we could reduce the dependency between the teams*".

Our survey results were aligned with the findings from the interviews that domain-driven design and bounded context patterns have become a mainstream in the CD context when designing CD-driven architectures. A majority of the survey respondents (54 out of 91, 59.3%) answered S20 (see Figure 11) as *strongly agree* or *agree*, while only 7.6% strongly disagreed or disagreed. Interestingly, this statement (S20) received the highest number of "*neutral*" responses (30 out of 91) by the survey respondents. Only 9 out of 30 respondents who selected "*neutral*" to respond to this statement introduced themselves as an architect. This may suggest that this statement is a highly specialized statement to architects, with which other roles may be unfamiliar. It is interesting to note that the majority of the interviewees and survey respondents believed that it was difficult to find the autonomous business capabilities of software architecture, as 81.3% of the survey respondents rated the severity of this challenge as *very important*, *important* or *moderately important* (see statement S6 in Figure 7).

**T2. Delay Decisions**: Software architecture in CD should be extremely adaptable to unpredictable and incremental changes [47]. That is why we found that it is difficult to make many upfront (architectural) design decisions in a CD context. Instead, the participants in our study only focused on an initial set of core decisions and other architectural decisions (e.g., decisions to choose a technology stack) are made as late as possible. Decisions are made when the time is right, for example, when requirements and facts are known [86]. This enables architects to keep architecture alternatives open to the last possible moment. Our findings show that deploying software changes frequently may necessitate making (architectural) design decisions on a daily basis. Furthermore, the participants also reported that the decisions were sometimes made unconsciously in a CD context. For example, **P16** stated: "*I try to avoid making big decisions. We probably need to decide what kind of technologies [are] useful for running microservices; but I try, even for those things, to keep them as flexible as possible. We probably want to change it [in the future] and I also try not to look down and make too many strong decisions*".

To support this claim further, empirically, we asked the survey participants how strongly they agree or disagree with this statement: "*compared with less frequent releases, we avoid big upfront architectural decisions for CD practice to support evolutionary changes*". As shown in Figure 11, 65% of the respondents (54 out of 83) strongly agreed or agreed that architectural decisions in a CD context are made as late as possible (see statement S17 in Figure 11).

**T3. Stateless Architecture**: Having a _stateless application_, in which there is no need to maintain the state within the application, was another technique used to support incremental changes at operations level, e.g., "*If a WAR file is completely stateless and if I want to deploy a new version, I can simply deploy a new version on the top of it and there is no state to keep for migrating to the new version*" *P10*.

**T4. Hide Unfinished Features at Run-time**: Our study shows that working in CD mode needs everybody to push their changes to the master branch on a continuous basis. We found from the interview study (i.e., confirmed by **P12**, **P20**, and **P21**) that having long-lived feature branches is a stumbling block for CD, as they are associated with challenges such as delayed feedback and increased merge complexity [87, 88]. The participants' organizations realized that instead of creating new branches, which developers may work on for a couple of months and then integrate back to the master, they need to switch off the long-lived feature (i.e., the unfinished feature) and then release it to end users only when the feature is ready. A _feature toggle_[4] pattern was indicated by some interviewees to achieve this goal. This pattern helps a team to release new changes to end users safely as the features that the team are developing are still in production, but nobody can see them because they are toggled. This pattern can be combined with and made more elegant by the _branch by abstract_[5] technique, which aims at making large-scale changes to a system in production. This technique first creates an abstraction layer around an old component and then gradually reroutes all interactions to the already-created abstraction layer. Once the new implementation of the old component finishes, all the interactions are rerouted to the new implementation. **P12** stated, "*We can also make feature toggle much more elegant by using branch by abstraction when you have implementation details separated out by interfaces and the child of implementation is based on toggles*". Other participants realized incremental changes by having side-by-side execution at component rather than at feature level. **P7** explained, "*We needed to improve the architecture to have side-by-side execution of components as it might be two versions of components, both installed at run-time. We could start to gradually cut the load over from the old version to the new version*".

> **Key Findings**
>
> *Finding 9.* Domain-driven design and bounded context patterns are increasingly used by organizations to design loosely coupled architectures based on autonomous business capabilities**.** However**,** most (81.3%) of the participants stated that identifying the autonomous business capabilities of software architecture is difficult.
>
> *Finding 10.* Compared with the less frequent releases, CD places greater emphasis on **evolutionary changes**. This requires delaying (architectural) design decisions to the last possible moment.

---

[4]https://martinfowler.com/articles/feature-toggles.html
[5]http://polysingularity.com/branch-by-abstraction/

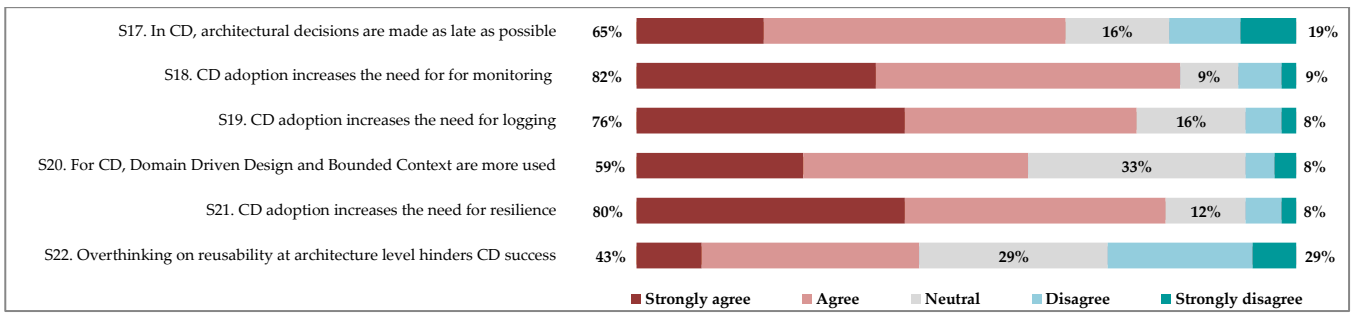

**Figure 11**. Survey responses to the statements on the quality attributes that need more attention in a CD context. While the importance of monitorability, loggability, and modifiability has increased, overthinking about "reusability" (confirmed by 43% of the surveyed participants) at architecture level may overturn CD adoption.

### 5.3.3 Loggability and Monitorability

Building and delivering an application on a continuous basis may potentially increase the number of errors, and the chance of unavailability of the application [1, 22]. This compelled the participants' organizations to have more investment in logging (i.e., the process of recording a time series of events of a system at runtime [1]) and monitoring (i.e., the process of checking the state of a system) in order to hypothesize and run experiments for examining different functionalities of a system, identify and resolve performance issues, and recover the application in case of problems. Therefore, the organizations practicing CD need to appropriately record, aggregate and analyze logs and metrics as an integral part of their CD environment. One interviewee highlighted this trend as "*We have everything based on the log. One aspect can be getting visibility in any part of the components, which are being pushed in the DevOps cycle. We already have it as part of our framework*" **P5**.

When we asked, "*since moving to CD practice, the need for monitoring (i.e., a centralized monitoring system) has increased*", the survey participants rated this statement (S18) *strongly agree* (36.2%), *agree* (46.1%), *neutral* (8.7%), *disagree* (6.5%), and *strongly disagree* (2.1%). As shown in Figure 11, the responses for statement S19 also indicate that 75.8% of the respondents strongly agreed or agreed that the need for a centralized logging system has increased since their clients or respective organizations moved to CD. In the following, we provide more insights into how monitorability and loggability are improved:

**T1. Use external tools**: Our analysis shows that there are <u>limitations in CD tools</u> (e.g., Jenkins), as they are not able to provide sufficient log data and also enable aggregation and analysis of logs from different sources. The participants' organizations extensively used external tools to address these limitations, e.g., "*So, basically those logs will be collected by other (external) systems. Then, we (the infrastructure team) can easily download and aggregate those logs on our own big data platform*" **P2**.

**T2. Build metrics and logging data into software** was another solution adopted by the participants' organizations. The participants stated that large-scale applications adopting CD need to expose different monitoring and logging data. For example, two interviewees mentioned: "*You build log capability such they can be turned on and off in the code in order to eliminate the performance problems*" **P6** and "*If we are talking about the logging aspect, if any tier fails for any particular reason, then an appropriate reason code must exist*" **P5**.

Regarding the two above-mentioned solutions, the interviewees raised two areas that need further consideration. Firstly, it is necessary to determine to <u>what extent the log data</u> should be produced (e.g., "*We have internal meeting discussions about how much effort we can put to capture [logs] per system; you know how much with the level of detail [is] appropriate*" **P7**). Secondly, <u>readability of the logs</u> for all stakeholders should be taken into consideration. As the developers mostly build the logging mechanism into the code, there is a chance that logs become unfathomable for operations teams and may not be efficiently employed to perform diagnostic activities: "*Support people (the operations team) feel that there is too much logging. So, they have to swim [in] too much information and find problems. The support people need to have cleaner and simpler logs to look at, where developers get all this information they need*" **P13**.

We also observed that the importance of loggability and monitorability increases considerably when a monolithic application is split into smaller units (e.g., microservices). This is due to the fact that splitting a monolith into multiple independently deployable services presents challenges in identifying problems (e.g., performance issues) in a system. Furthermore, it is needed to trace how services go through a system. It is essential to ensure that there is enough trace information for logs.

**Key Findings**

*Finding 11. While software organizations **extensively use monitoring tools** to address the shortcomings of CD tools, the applications in a CD context should **expose the different types of log and monitoring data with** a standard format. This helps build architectures that are responsive to operational needs over time.*

### 5.3.4 Testability

One of the main concerns stated by the participants in transition to CD was about *testing*. It was observed during the interviews that addressing testability concerns influences architectural design decisions for explicitly <u>building testability inside a system</u>. It is emphasized that architectures in a CD context should support testing to the extent that tests should be performed quickly and automatically. Another solution adopted by the participants to improve testability was using the right tools and technologies. Nevertheless, the participants worried about the potential limitations of tools, as tools alone are insufficient for addressing all the testing problems. For example, we were told that existing automated testing tools tend to be fragile in some types of applications (e.g., mobile applications) or environments (see [67] for more information). That is why the participants typically approached the testing challenges by combining architectural (i.e., testability inside the design) and technological (i.e., selecting the right tools) solutions. Apart from choosing appropriate testing tools, we found the following techniques support testability inside a system:

**T1. Improving test quality**: To avoid putting overhead on a CD pipeline, it was frequently suggested that there should be improvements to the quality of tests (data) rather than increasing the number of test cases. One interviewee put it, "*We are looking at the quality to improve the DevOps model. One of the things that we can improve is the quality of the tests themselves; do we have the right test suites? Do we have the right coverage?*" **P6**. This includes improving quality by selecting the right number of test cases to be performed at the right time on quality test data (it was also affirmed by **P5**, **P12**, **P15**, and **P17**). The cycle time of a CD pipeline (i.e., the required time to push code from the repository to the production environment) can be increased by long-running tests. Furthermore, long-running tests can slow down the feedback cycle in a CD pipeline. Respondent **R5** commented that running tests in parallel can decrease the cycle time of a CD pipeline. These issues happen when test cases are poorly written and designed. One of the interviewees (**P12**) described how improving the quality of tests could help to accelerate the feedback cycle in the CD pipeline. He highlighted two practices: (i) designing smaller test units that result in immediate feedback, (ii) decreasing the number of functional tests. The applied practices resulted in receiving the same level of feedback from applications, in a timely manner.

**T2. Making code more testable**: It was also found that testability should be addressed at code level. Whilst services and components need to be tested independently, testing them within a larger CD pipeline with all dependencies can be a complicated and costly process. For instance, it may be costly to provide all the resources (e.g., databases) for testing in staging environments. A typical solution to this problem was dependency injection, which would bring more flexibility to the testing process. One participant opined: "*From the architectural standpoint, we very often build projects that use dependency injection mechanism in, like Spring, so that we can inject a set of dependencies. For example, you inject an in-memory fake database rather than a production database, you inject a fake set of resources rather than relying on [real] resources, which makes the code much more easily testable*" **P13**.

Another example comes from participant **P6**, who built performance tests into the code. Rather than just running performance tests as black box testing from outside of the code, performance testing was actually tightly integrated into the code. To put it another way, the code itself would be measuring start and stop times between operations and between events occurring within the code. All these times (i.e., the start and stop times for different operations in the system) are recorded to compare them with a baseline to decide whether there is a problem or not. According to **P6**, this practice significantly improved the overall performance of the system.

**Key Findings**

*Finding 12. CD practices require **high quality** tests that **run easily**. Therefore, it is necessary to make informed (architectural) design decisions to improve the testability of a system by (1) improving test quality (e.g., more suitable, simpler test methods and decreasing test cycle times); (2) making code more testable; (3) using smaller components.*

### 5.3.5 Resilience

Increasing the frequency of deployment to production may increase the number and severity of failures. So, failures are inevitable in a CD context. Through the interview study, we found that for some participants, designing for failure guides their architecture designs. **P11** characterized the software architecture in a CD context as follows: "*an architecture that splits the system also by units that should fail independently*" and **P14** told us, "*resilience is also a matter of integration aspect: how to integrate with the* *third party [units]; what if a third party starts to fail*". We also observed that many discussions among the participants were centered on resilience versus reliability as the main concern was *not* to prevent failures but to think about how fast the failures can be identified and isolated, and how to automatically rollback to an early, stable version. An example of this trend is vividly typified in the following anecdote: "*I think when we are moving to this paradigm [CD], rather than ensuring your software is gold before you deploy it to production, for example, by three month testing cycles to validate every bit of it, we are going to a mode [where the] mean time to recovery [is important]. [For example] if I face an issue in production, how soon I can recover from that failure*" **P10**.

Whilst we could not conclude that the resilience quality attribute needs to be prioritized over reliability within a CD context [89], the responses to statement S21 indicate that the survey participants are significantly concerned with resilience when designing and implementing applications. This quality attribute appeals to a large portion of respondents: 80.2% strongly agreed or agreed that the CD practices increase the need for resilience (see statement S21 in Figure 11).

> **Key Findings**
>
> *Finding 13. Design for failure is considered as the foundation of the architecture in a CD context, in which, instead of preventing failures (reliability), it is more important to learn how to deal with failures (resilience).*

### 5.3.6 Reusability

Our analysis reveals the architecture designs that emphasize reusability could make practicing CD more challenging. The drawbacks of overthinking on reusability at architecture level (e.g., using shared components, or packaged software) in the context of CD are two-fold: **(1)** *inter-dependencies between software development teams* increase in the sense that they rely on shared software units. It means that changing the shared units requires seeking inputs and reaching agreements among all the relevant stakeholders, which demands significant time and effort (e.g., "*If you really do not control explicitly what you really expose and what you really reuse, you will end up with a lot of mess in the code.*" **P14**). This approach hinders the *autonomy of software teams* in building and deploying software components and applications. **(2)** The other concern is about being able to *test the configuration of shared software units* (e.g., well-known packaged software). Our interviews revealed that it is vital for the application developers working in the context of CD to write test cases, through which they can validate the configuration of software packages used in the application. Hence, reusing packaged software may hinder their ability to perform potential unit tests and push the packaged software into a CD pipeline on a continuous basis. This leads to *losing the testability of those configurations*, which is significant. As **P14** explained: "*We had plenty of products and there are a lot of [common] things between them. We reused [things] like utilities and even some part of the domain. We spent a vast amount of time maintaining agreements between teams; [for example] how to evolve them. We decided to fork them; just go your own way, whatever you want and just leave us alone, because reuse has side effects. So, we have to understand duplication is not evil*".

As shown in Figure 11, about 42.8% of the survey respondents (39 out of 91) strongly agreed or agreed that overthinking about *reusability* at architecture level (e.g., reusing packaged software) hinders CD success, while 28.5% disagreed or strongly disagreed. This fairly agrees with the interview findings. We found that such attitudes were mostly dominant among software architects, as 20 out of the 39 architects who participated in the survey have voted statement S22 as *strongly agree* and *agree*, while only 7 architects rated it as *strongly disagree* or *disagree*. Other respondents (28.5 %) had no idea whether or not having too much reusability is a major roadblock for CD.

> **Key Findings**
>
> *Finding 14. Overthinking about reusability at architecture level has two side effects: (1) increased inter-team dependency and (2) losing testability. However, among other roles (e.g., developers), software architects strongly believe that a lack of explicit control on reuse at architecture level (e.g., shared components) would make practicing CD harder.*

## 5.4 Perspectives on Operational Aspects

The importance of operational aspect management (i.e., of the production environment, operations stakeholders and their requirements) for the architecting process in a CD context was highlighted during the interview study. We observed that the participants' organizations have been shifting from considering operational aspects as separate and sometimes uncontrolled entities to treating them as first-class entities in the architecting process, particularly after CD adoption [1]. The shift from the first to the second paradigm was also confirmed by the survey results, in which 83.5% of the respondents indicated that they consider operational aspects during the architecting process *almost always* or *often* (see statement S16 in Figure 10). Additionally, the respondents believed that operations requirements and aspects significantly impact on architecture design and design decisions (see Figure 5). 83.5% of the surveyed practitioners strongly agreed or agreed with statement S24 in Figure 12 ("*Operational aspects and concerns impact on our architecture design decisions*"). Other studies have also found that involving relevant stakeholders in the architecting process leads to informed and balanced architectural decisions [90, 91]. This might help to mature the architecting process [82]. In the following subsections, we provide more insights into why operational aspects need to be perceived as important in a CD context.

### 5.4.1 Production Environment Challenges

The interview study revealed that production environments may pose challenges to architecture design. One of the top challenges is to make sure software changes are being seamlessly transferred and deployed into multiple, <u>inconsistent</u> production environments. A program manager (**P4**) described the challenge of inconsistent environments thus "*One challenge is that there are multiple environments. [We must] make sure that the system of development is compatible with all environments, which means we [need to] give the current infrastructure, hardware and technical details of all environments*". <u>Regulatory and controlled environments</u> were also stumbling blocks to CD as these environments usually follow a formal deployment process. There is a need to adjust and adapt architectures in the regulated environments so that apart from those parts of a system that are not really able to adopt CD, the rest are being deployed on a continuous basis. As **P14** mentioned, "*Architecture is also about compliance and security as well. Because let's say there is a kind of regulated environment and you want to have CD there; how [should we] adjust the architecture? For example, you can rip out part of your software and make sure it is part of your architecture which can't really support CD but let the rest of the architecture and software be deployed in a CD fashion*". Our previous work [67] shows that a lack of control of the production environment by the development team negatively impacts on deployment frequency (e.g., "*Inside the client network, the processes are less mature and more overhead exists to deploy code to controlled environments, which requires manual processes to turn over code*" **P3**).

Another category of the challenges is related to <u>the extent to which operations requirements</u> need to be collected and shared with software architects. Taking into account too many requirements in an operations environment could result in designing a system too specific for that environment. This introduces the risk that design decisions become dependent on the operational configurations and are therefore fragile to the changes that are made in that environment. On the other hand, ignoring the characteristics of an operations environment during the architectural design of an application is likely to result in losing the operability of the designed architecture. This challenge was explained by **P5**, "*They [the operations team] had purchased some infrastructure and some hardware and what they wanted from us [was] that our solution (architecture design) should be in compliance with their already purchased infrastructures. This all added to the challenges [to the architecture design]*".

All the above challenges may lead to the expansion of the role of architects. Architecting for CD goes beyond designing a software application: it requires incorporating consideration of the whole environment (e.g., database, automation, and test data). Architects are expected to look into a broader spectrum and propose solutions that keep consistency after the application is deployed and upgraded. This implies versioning the whole environment rather than versioning the source code alone (i.e., everything-as-code [92]). One interviewee described this trend in these words: "*It is not architecting things in terms of source code, but it is more about architecting the deployment as a larger entity, [for example], in finding ways to drive and maintain the consistency across those environments after we deploy and upgrade because they are on different servers. It is not [similar to] a cloud system in a multi-tenant environment*" **P7**.

### 5.4.2 Operations Stakeholders and Requirements

In this section, we discuss what strategies have been applied in industry to treat the operations team and their concerns as the first-class entity in the software development process, which also affects the architecting process (see Figure 5).

**S1. Leveraging logs and metric data in an appropriate format for operational tasks**: The most cited strategy, indicated by 62 (68.1%) survey respondents, refers to collecting and structuring logs, metrics (e.g., CPU and memory usage) and operational data in appropriate formats to enable the operations team to make faster and more informed operational decisions in post-deployment activities such as assessing the deployed application in production. For example, we have: "*One of the key aspects [of CD] would be around having sufficient logging in place, which you can identify when things go bad. You have to build a dashboard to tell you what the current load is and also aggregate that data on the period of time*" **P12**.

**S2. Early and continuous engagement of operations staff in decision-making**: The participants also reported that early and continuous involvement of the operations team in the development process, particularly in the design process, boosted the productivity of their respective organizations in delivering value on a continuous basis. It is mainly because the deployability of the application is more thoroughly considered in the development and design phases; the deployment process would not be a big issue at the end. As shown in Table 2, this was the second most cited (55, 60.4%) strategy by the survey participants. Here are just a few of the example quotations supporting this strategy:

"*Once the operations team was brought into planning meetings, most of the operations requirements could be negotiated off and we would find cheaper wins that would be less expensive than the original plans of the software developers. So those operations requirements could be better implemented and landed in production*" **P17**.

**S3. Prioritizing operations concerns**: It is argued that the operations team has a set of concerns (e.g., quickly detecting and predicting failures), which should be effectively addressed [1]. During the interviews, we learned that despite the adoption of CD, operations concerns and requirements still have a lower priority than other requirements. This is moderately confirmed by the survey study as 41.7% of the survey respondents (strongly) agreed with this finding (see statement S23 in Figure 12). It is interesting to note that there is a relatively large proportion of the respondents (29.6%) who are neutral about statement S23 and only 28.5% strongly disagreed or disagreed. The interviewees revealed several reasons for deprioritizing operations requirements. For example, developers usually believe that the role of the operations team is only to serve the development team. It has been discussed by the interviewees that people who historically decide on priorities (e.g., the product owner) usually do not consider so much business value from operational aspects. According to the interviewees, <u>ignoring and deprioritizing operations requirements</u> has resulted in severe bottlenecks for deploying applications on a continuous basis. The interviewees' organizations started treating the operations personnel and their requirements as having the same priority as others. So, the operations requirements became an integrated part of the requirements engineering and architecture design phases. Table 2 shows that 41 (45.1%) of the surveyed practitioners adopted this stagey. **P18** referred to this strategy as follows: *"One thing [that] we are doing for this client, for this software is to emphasize the importance and address operational things much more. That is starting to happen, but it is quite a big change"*.

When we asked the participants about their experiences in thinking about operational aspects from day one of a project, we noted several benefits: (i) it puts operations staff on the team and reviews their concerns quite early; (ii) it influences how developers look at applications in production effectively; (iii) it forces architects to think much more about the problems that may happen in the deployment process (e.g., "*Only recently we have realized that operations concerns are a really fundamental part of architecture. If you take that as part of the architecture and treat them as first-class citizens, it allows you to build an architecture that is much more flexible*" *P13*); (iv) it provides much more flexibility and enables organizations to get applications deployed much faster and therefore realize CD.

We also added the "Other" field to gather more strategies in this regard. Interestingly, only 8.7% of the survey respondents filled in the "Other" field. The respondents mostly used the "Other" field to elaborate their previous choices (e.g., "*Ensuring we have scheduled placeholders to revisit changes in operational requirements, and that we have an operational requirements champion*" *R16*, "*Operations [team] owns the final solution and must sign off on it*" *R40*).

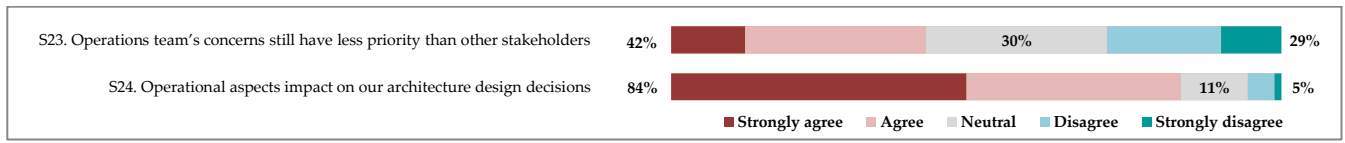

**Figure 12**. Survey responses to the statements on operational aspects.

**Table 2**. Strategies for increasing the amount of attention paid to operations team and their concerns

| Strategy | # | % |
|---|---|---|
| Leveraging logs and metric data in appropriate formats for operational tasks | 62 | 68.1 |
| Early and continuous engagement of operations staff in decision-making | 55 | 60.4 |
| Prioritizing operational concerns | 41 | 45.1 |
| Others | 8 | 8.7 |

---

**Key Findings**

*Finding 15.* The requirement for continuously and automatically deploying to unpredictable production environments can **expand the role of architects** as, apart from software design, they need to collaborate substantially with operations personnel and deal with infrastructure architecture, test architecture, and automation.

*Finding 16.* Designing highly operations-friendly architectures are achieved by (1) leveraging logs and metric data; (2) engaging operations personnel in decision-making; (3) prioritizing operational concerns; and (4) understanding production environments' constraints (i.e., constraints in **inconsistent**, **regulatory** and **controlled** environments).

---

## 6. Discussion and Conclusion

This work has empirically explored how an application should be (re-) architected for CD. We applied a mix-methods approach consisting of semi-structured interviews and an online survey for collecting qualitative and quantitative data, which have been systematically and rigorously analyzed using appropriate data analysis methods. The findings from this study are expected to make significant contributions to the growing body of evidential knowledge about the state of the art of architecting for CD practices. Furthermore, the results of this research can provide improvements to the adoption of DevOps and CD. In this section, we first discuss this study's main findings while comparing and contrasting them with the most related literature on this topic. Then, we draw some implications of this study for practitioners. It is worth noting that from the methodological perspective, our study includes varied demographics rather than the peculiarities of practitioners' own experiences [1, 22, 42, 46] or a single case company and particular domain [22, 45].

## 6.1 Main Findings

**Monoliths and CD are not intrinsically oxymoronic.** Our conclusion is in line with the results of [19, 24, 46] that monoliths present significant challenges to CD success. Yet, we have found some examples in both the interview and the survey parts of this study that it is *possible* to do CD with monoliths. This has also been indicated by personal experience reports by Prewer [93], Vishal [21], Schauenberg [94] and Savor et al. [38]. However, an overwhelming majority of the interviewees and the survey participants believe that it is much more difficult to adopt CD within large monolithic systems or large components. Compared with the work by Bass et al. [1], Schermann et al. [24], and Newman [46], our study has independently identified the reasons why practicing CD within monoliths is difficult (**Finding 1**). We found that growing monoliths leads to increased complexity of internal and external dependencies, restricted automation (test and deployment), impeding teams' ownership and having slow and inconsistent feedback [19, 24, 73], which together can be roadblocks to frequent and automatic deployment. Finally, we have provided principles, practices, and strategies with concrete examples to achieve CD with monoliths. Specifically, those organizations that effectively embraced CD for their monoliths mostly employ the following practices to optimize their deployments: (i) developing highly customized tools and infrastructures for monitoring and logging [21]; (ii) reducing test run times by improving the test quality (e.g., parallelization of automated testing); (iii) describing, testing and deploying components/services *only* through interface specification with backwards capability; and (iv) deploying all entities including components and databases through a rigorous CD pipeline.

**Characteristics of the principle of "small and independent deployment units".** All of the above-mentioned monolith-related challenges compelled many participants' organizations to move beyond the monoliths to facilitate the CD adoption journey. Similar to Newman [46], and Lewis and Fowler [48], we can conclude that the principle of "small and independent deployment units" is an alternative to monolithic systems for this purpose and serves as a foundation to design CD-driven architectures (**Finding 4**). We observed that this principle has mostly been implemented in the industry using two concrete architectural styles: vertical decomposing and microservices. Our findings particularly emphasize that adopting these architectural styles may result in overhead costs to deal with complex deployment processes, which require having sophisticated logging and monitoring mechanisms and high operations skills (**Finding 6**) [83]. For example, the microservices style requires every team to build a CD pipeline for each component or service rather than having one CD pipeline for the whole application. However, recently, tools such as Spinnaker[6] have been developed which help teams to have global control of various pipelines with different stages, quality control processes, and release cadences. Software organizations incrementally split their applications into "small and independent deployment units" that can be independently managed. Our study includes other scopes compared with [46, 48], as we also focused on the characterization of "small and independent deployment units" (**Finding 5**). We have proposed that "deployment units" need to represent *only* one business domain problem (i.e., should not cross its bounded context) and each of the "deployment units" should be autonomous in terms of *deployability*, *modifiability*, *testability*, and *scalability*. These factors are among the highest rated statements by the participants (i.e., more than 68% of the participants (strongly) agreed with all these factors). That means organizations can use these factors as reliable criteria when moving towards CD. Other factors such as having team-scale autonomy per "deployment unit" are desirable but less cited by the participants.

**First class quality attributes in the CD context.** Our findings indicate that CD and its associated constraints (e.g., unpredictable environments) expect a certain kind of architecture to emerge. That means CD practices significantly change the priority of some quality attributes such as deployability, modifiability, testability, monitorability, loggability, and resilience. Whilst these quality attributes are important for all types of contexts and systems, they are considered critically important for a CD-driven architecture design.

Previous research also highlights the importance of these quality attributes [22, 23]. Bellomo et al. [23] only focus on deployability as a new quality attribute in three projects and introduce tactics to achieve deployability. In contrast to [23], we have investigated the impact of deployability on different aspects of (architecture) design (high-level vs. low-level design) and also we indicated that deployability has a minimum conflict with other quality attributes (**Finding 8**). Chen [22] also highlights what quality attributes (i.e., deployability, modifiability, security, monitorability, testability, and loggability) can be more important in the CD context; however, he did not provide any details about how these quality attributes can impact on the architecture design and how they can be achieved. Our results confirm and extend the previous findings [22, 23] (i) by improving our understanding of how the aforementioned quality attributes affect architectural decisions; and (ii) by introducing techniques to achieve these quality attributes (e.g., keeping components stateless helps improve modifiability) (**Findings 8 to 14**). Whilst we targeted the quality attributes of applications in the CD context, Bass et al. [1] focus on the quality concerns of a CD pipeline rather than applications. They indicate that the primary quality concerns that need to be built early in a CD pipeline include repeatability, performance, reliability, recoverability, interoperability, testability, modifiability, and security.

---

[6]https://www.spinnaker.io/

Furthermore, we have observed that from an architect's perspective, "reusability" at the architecture level is the only quality attribute that should *not* be overthought, as it contributes to overturning CD adoption (**Finding 14**). It should be noted that we could not conclude that CD substantially influences other quality attributes (e.g., security as indicated by Chen [22]). On the other hand, our study highlights that CD practices seek software architectures that are easily composable, self-contained, and less monolithic with less shared data among components. Architectures in CD should be atomized for autonomy and independent evolution to make sure there is a minimum overlap between teams (**Finding 2**). Consequently, software architects should avoid upfront (architectural) design decisions and delay them to the last possible moment to effectively meet the aforementioned quality attributes (**Finding 10**).

**Operations-friendly architectures**. One of the most intriguing findings of this study is that there is an increasing trend of paying significant attention to operational aspects (e.g., operations requirements) in the architecting process. Our observation from our previous work [67] and this study is that the constraints of the production environment (e.g., lack of access to the production environment) may slow down the pace of production deployment. We can conclude that one of the reasons for not continuously and automatically deploying into production environments stems from not carefully understanding and dealing with the production environments' constraints. We recommend that software development organizations invest time in understanding their production environments and their respective constraints that may not support CD. Accordingly, software architects need to collaborate substantially with operations teams during the requirements engineering and architecture design phases (**Finding 15**). Though we know from other studies (e.g., Bass et al. [1]) that operations teams and their requirements play significant roles in CD adoption, our study is the (first) large-scale piece of empirical work that has specifically explored the strategies (e.g., early and continuous engagement of the operations team in architecture design) that are practiced in industry to achieve this goal (**Finding 16**). We emphasize that considering the operational aspects of a system early in the software development process helps to scale the complexity of the deployment process at approximately the same pace as the code base. This trend has changed the role and responsibilities of the architect (**Finding 15**). From the research perspective, further investigation is necessary for understanding how software architects perform their work in a CD context to provide guidance for the required job skills and education programs [3, 28].

## 6.2 Implications

It is argued that the findings of empirical studies should provide insights for practitioners. At the end of the online survey, the participants were asked if they wanted to receive the results of our study by providing their email addresses. Surprisingly, 81.3% (74 out of 91) of the respondents provided their email as they were interested in receiving the findings of this study. We believe this indicates that our findings are highly likely to draw the interest of software engineering practitioners. Based on the findings from this study, we draw some implications for research and practice.

**Deployable units do not only mean software components/services**. The results of our study suggest the concept of "*Continuously Deployable Units*". This indicates that in addition to software components/services, every entity (e.g., database and dependencies) that an application depends on, should be a unit of deployment in a CD pipeline. We have observed that the database should be continuously provided with as much automation support as possible for successfully implementing CD. Whilst the participants in our study employed some practices (e.g., schema-less databases) and used appropriate tools for this purpose (**Finding 7**), it is clear that there is not much support for automatically delivering database changes on a continuous basis [15, 95]. Our study reveals that there are a number of organizations that could not succeed with CD due to a lack of database automation in their CD pipeline. The reason for this stems from the fact that validating data and managing schema migration in the CD pipeline remain immature (e.g., manual steps requiring DBA involvement). Hence, organizations prefer to do it offline. We believe that this can be a reason that our study's participants rated the database-related challenges as the least important in their CD journey (see Section 5.1.1). On the research side, more focus is needed on mining and analyzing database-related failures in the CD pipeline to thoroughly understand the behavior of databases in a CD context. This helps to design effective and robust solutions (e.g., tools) to support organizations in automating the deployment of schema modifications and data conversions in CD pipelines without compromising the continuous deployment of application code changes [15].

**Adjust the organizational structure to support CD**. Our experience in this study was that the process of changing to a CD organization requires more than providing tool support and automation. CD impacts on organizational structures (e.g., team structures), roles and responsibilities. From practitioners' perspectives, the inflexibility of organizational structures with the spirit of CD practices is the most critical challenge for implementing CD (**Finding 3**). A software organization can succeed in

Preprint - accepted to be published in Empirical Software Engineering (2018)

CD adoption once there is flexibility in optimizing organizational structures to be aligned with CD and certain skills and responsibilities need to be sought. Finding the best organizational structures that suit an organization depends on many factors, such as the flexibility of the current organizational structure, available skills and management procedures [96, 97]. Answering this question requires extensive empirical studies to explore these organizational structures, supporting CD practices in different situations. Our previous study [98] has provided preliminary findings about how team structures, roles, and responsibilities can change while adopting CD. We have identified four main patterns for organizing development and operations teams (e.g., no visible operations team) in industry. We believe that there is a need for further research to gain an in-depth understanding of suitable organizational structures for implementing CD.

**Real-time, digestible and customizable monitoring is key**. As elaborated in Section 5.3.3, we have seen the growing importance of monitorability and loggability in the transition to CD. In fact, there is a growing tendency to adopt log-driven, log-specific architectures that support the continuous collection of operational data, facilitate aggregation of logs and transform them into a searchable format [99]. More importantly, breaking down a monolith into "small and independent deployment units" can worsen the monitoring challenges because tracing a large number of independently deployable units and identifying problems inside a system can be even more challenging. Such issues can significantly impact the way that a system is composed of smaller pieces. Whilst a number of solutions (e.g., *extensive use of monitoring tools*, *creating monitoring and log data into applications*) have been introduced, there are still several challenges to be addressed in order to implement monitoring and logging in a CD context. Based on our knowledge, two main areas that usually cause practical challenges in this regard are the *readability of logs for all stakeholders* and *the abundance of logs and monitoring data*. We believe these challenges can be alleviated by (1) establishing a standard for logging and monitoring by which all developers and applications in an organization create consistent logging and monitoring data [100], and (2) applying analytical techniques (e.g., machine learning approaches) to summarize, prioritize or filter monitoring and logging data. Our findings re-emphasize the conclusion by Bass et al. [1] and Humble and Farley [42] that to achieve effective monitoring in the CD context, there is a need to build a centralized platform for providing real-time, digestible and customizable monitoring and alerting for the different types of stakeholders. This would enable stakeholders to understand what happens and why in real (or near) time.

# 7. Acknowledgment

The authors would like to thank all participants in this study. This work is partially supported by Data61, a business unit of CSIRO, Australia. The first author is also supported by Australian Government Research Training Program Scholarship. We also greatly appreciate the hard work and time spent by the anonymous reviewers and the handling editor in providing insightful comments and help us to improve the manuscript.

# 8. Appendix A. Interview Guide

**Interviewee's background**
**Q1**. What are your role and responsibilities in the project team?
**Q2**. How long have you performed that role? Are you an architect?
**Q3**. Please talk about the size and domain of your organization.
**Project's description**
**Q4**. What is/was the domain and type of project?
**Q5**. Team size: how many people are/were involved in this project?
**Q6**. How often is/was the application in a releasable state? How often do/did you deploy the application to production or the customer?
**Operations stakeholders and their requirements**
**Q7**. How do/did you consider operations requirements in the design process?
**Q8**. Do/Did you involve operations stakeholders in the decision-making process at the early stage of software development? If so, how?
**Q9**. Do/Did you have any difficulties in prioritizing the concerns of operations stakeholders with other stakeholders? How did you manage them?
**Architecture and Continuous Delivery/Deployment Practices**
**Q10**. How would you describe the relationship between architecture and CD practice? How does the adoption of CD practice change the architecting process?
**Q11**. How do/did you design the architecture of your system to enable and better support CD practice e.g., improving deployability?
    **Q11.1** How do/did you predict and evaluate the deployability of your software system during the design process?
    **Q11.2** Can you provide one example of architectural decisions you made for improving deployability?
**Q12**. What architecture principles and practices (e.g., patterns, styles, and tactics) did you employ to promote and support CD practice?
    **Q12.1** How do you break down the monolithic applications into independently deployable units/components/services?
    **Q12.2** Are you using any criteria for this purpose?
    **Q12.3** Are you using the microservices style for this purpose? What about other techniques for this purpose?
**Q13**. What challenges do/did you experience whilst designing the application architecture for CD?

**Q14**. What quality attributes are more influenced by the CD context? How is this so?
**Q15**. What quality attributes are in support of or in conflict with deployability? How is this so?

## 9. Appendix B. Survey Questions

| Questions | Scale |
|---|---|
| **Demographic Questions** | |
| How many years have you worked in software or IT industry? | 0–2 / 3–5 / 6–10 / > 10 years |
| What is your role in the development project? | Developer / Architect / Tester / QA / … |
| How large is your organization? | 1–100 / 101-1000 / >1000 employees |
| What is the domain of your organization? | Consulting and IT services / Embedded system/ … |
| **Practicing Continuous Delivery and Continuous Deployment** | |
| On average, how often is your application in a releasable state (i.e., production-ready)? | Multiple times a day / Once a day / A few times (e.g., one or two) a week / A few times (e.g., one or two) a month / A few times (e.g., one or two) a year |
| On average, how often do you deploy/release an application to production or the customer environment? | Multiple times a day / Once a day / A few times (e.g., one or two) a week / A few times (e.g., one or two) a month / A few times (e.g., one or two) a year |
| **Operations Stakeholders and their Concerns** | |
| Despite the adoption of CD in my organization, the operations team's concerns and requirements still have a lower priority than other stakeholders. | Strongly agree / Agree / Neutral / Disagree / Strongly disagree |
| To increase the amount of attention paid to operations team and their concerns, my organization has adopted the following strategies: | Prioritizing operations concerns (i.e., consider the operations and their requirements as being as important as others) … |
| | Early and continuous engagement of Ops staff in the decision-making process for the development process (i.e., design process). |
| | Leveraging logs and metric data for operational activities. We collect and structure logs, metrics (e.g., CPU usage) and operational data in appropriate formats to enable … |
| | Other (Specify): |
| **Software Architecture and Quality Attributes in Continuous Delivery and Deployment** | |
| How would you grade the importance of software architecture design in successfully adopting and implementing CD practice? | Very important / Important / Moderately important / Of little importance / Unimportant |
| When we are designing the architecture of an application, we also consider the operational aspects, requirements and concerns (e.g., to make the architecture readily supportive of CD). | Almost Always / Often / Sometimes / Rarely / Never |
| Operational aspects and concerns impact on our architecture design decisions. | Strongly agree / Agree / Neutral / Disagree / Strongly disagree |
| It is "possible" to successfully practice CD in "monolithic applications". | Strongly agree / Agree / Neutral / Disagree / Strongly disagree |
| **In order to break down (monolithic) applications into smaller and independent units/components/services as STRONGLY recommended by CD practice, how would you define small services/components/units in your organization?** | |
| A component/service is small if it can be scaled independently. | Strongly agree / Agree / Neutral / Disagree / Strongly disagree |
| A component/service is small if it can be deployed independently. | Strongly agree / Agree / Neutral / Disagree / Strongly disagree |
| A component/service is small if it can be tested independently. | Strongly agree / Agree / Neutral / Disagree / Strongly disagree |
| A component/service is small if it can be modified independently. | Strongly agree / Agree / Neutral / Disagree / Strongly disagree |
| Can you describe (e.g., in one sentence) how a component or service should be to be suitable for successfully practicing CD? | Free text |
| In the projects that have adopted or are adopting CD practice, deployability concerns impact(ed) the design of individual classes. | Almost Always / Often / Sometimes / Rarely / Never |
| In the projects that have adopted or are adopting CD practice, deployability concerns impact(ed) the design of individual components/services. | Almost Always / Often / Sometimes / Rarely / Never |
| In the projects that have adopted or are adopting CD practice, deployability concerns impact(ed) the design of interactions among components/services. | Almost Always / Often / Sometimes / Rarely / Never |
| In the projects that have adopted or are adopting CD practice, deployability concerns impact(ed) the design of an entire application. | Almost Always / Often / Sometimes / Rarely / Never |
| In order to improve deployability of an application, I can sacrifice performance, security, usability, etc. | Almost Always / Often / Sometimes / Rarely / Never |
| Focusing too much on reusability at component or application level can be a bottleneck to continuously deploying software. | Strongly agree / Agree / Neutral / Disagree / Strongly disagree |
| Since moving to CD practice, the need for monitoring (i.e., having a centralized monitoring system) has increased. | Strongly agree / Agree / Neutral / Disagree / Strongly disagree |
| Since moving to CD practice, the need for logging (i.e., having a centralized logging system) has increased. | Strongly agree / Agree / Neutral / Disagree / Strongly disagree |
| Since moving to CD practice, Domain Driven Design and Bounded Context patterns have been applied MORE often and practiced when designing loosely coupled architectures. | Strongly agree / Agree / Neutral / Disagree / Strongly disagree |
| Compared with less frequent releases, we avoid big upfront architectural decisions for CD practices to support evolutionary changes (i.e., architectural decisions are made as late as possible). | Strongly agree / Agree / Neutral / Disagree / Strongly disagree |
| CD practice increases the need for resilience (i.e., design for failure). | Strongly agree / Agree / Neutral / Disagree / Strongly disagree |
| **Challenges and Barriers to Adopting Continuous Delivery and Deployment Practices** | |
| How important are the following challenges (if any) during adopting and implementing CD and which may put you in trouble? | |
| Huge dependencies and coordination among software development team members. | Very important / Important / Moderately important / Of little importance / Unimportant |
| Difficulty of splitting a (monolithic) application into independently deployable and autonomous components/services. | Very important / Important / Moderately important / Of little importance / Unimportant |
| Inflexibility of the organization's structure with the spirit of CD practice. | Very important / Important / Moderately important / Of little importance / Unimportant |
| Difficulty of breaking down a single-monolithic database into smaller and continuously deployable databases (i.e., decentralized data). | Very important / Important / Moderately important / Of little importance / Unimportant |
| Difficulty of identifying autonomous business capabilities. | Very important / Important / Moderately important / Of little importance / Unimportant |

[54] L. A. Palinkas, S. M. Horwitz, C. A. Green, J. P. Wisdom, N. Duan, and K. Hoagwood, "Purposeful Sampling for Qualitative Data Collection and Analysis in Mixed Method Implementation Research," *Administration and Policy in Mental Health and Mental Health Services Research,* journal article vol. 42, no. 5, pp. 533-544, 2015.
[55] L. A. Goodman, "Snowball Sampling," *Annals of Mathematical Statistics,* vol. 32, no. 1, pp. 148-170, 1961.
[56] D. S. Cruzes and T. Dyba, "Recommended Steps for Thematic Synthesis in Software Engineering," in *International Symposium on Empirical Software Engineering and Measurement (ESEM)*, 2011, pp. 275-284.
[57] C. B. Seaman, "Qualitative methods in empirical studies of software engineering," *IEEE Transactions on Software Engineering,* vol. 25, no. 4, pp. 557-572, 1999.
[58] V. Braun and V. Clarke, "Using thematic analysis in psychology," *Qualitative research in psychology,* vol. 3, no. 2, pp. 77-101, 2006.
[59] G. Gousios, M.-A. Storey, and A. Bacchelli, "Work practices and challenges in pull-based development: the contributor's perspective," in *38th International Conference on Software Engineering*, Austin, Texas, 2016, pp. 285-296: ACM.
[60] E. Murphy-Hill, T. Zimmermann, C. Bird, and N. Nagappan, "The Design Space of Bug Fixes and How Developers Navigate It," *IEEE Transactions on Software Engineering,* vol. 41, no. 1, pp. 65-81, 2015.
[61] A. W. Meade and S. B. Craig, "Identifying careless responses in survey data," *Psychological methods,* vol. 17, no. 3, p. 437, 2012.
[62] B. Kitchenham, L. Pickard, and S. L. Pfleeger, "Case studies for method and tool evaluation," *IEEE Software,* vol. 12, no. 4, pp. 52-62, 1995.
[63] J. Cito, P. Leitner, T. Fritz, and H. C. Gall, "The making of cloud applications: an empirical study on software development for the cloud," in *10th Joint Meeting on Foundations of Software Engineering*, Bergamo, Italy, 2015, pp. 393-403: ACM.
[64] F. Adrian, "Response bias, social desirability and dissimulation," *Personality and Individual Differences,* vol. 7, no. 3, pp. 385-400, 1996.
[65] F. J. Fowler Jr, *Survey research methods*. Sage publications, 2013.
[66] L. I. Meho, "E-mail interviewing in qualitative research: A methodological discussion: Research Articles," *J. Am. Soc. Inf. Sci. Technol.,* vol. 57, no. 10, pp. 1284-1295, 2006.
[67] M. Shahin, M. A. Babar, M. Zahedi, and L. Zhu, "Beyond Continuous Delivery: An Empirical Investigation of Continuous Deployment Challenges," in *11th ACM/IEEE International Symposium on Empirical Software Engineering and Measurement (ESEM)*, Toronto, Canada, 2017: IEEE.
[68] P. J. Andre van Hoorn, Philipp Leitner, Ingo Weber, "Report from GI-Dagstuhl Seminar 16394: Software Performance Engineering in the DevOps World. Available at: https://arxiv.org/abs/1709.08951," 2017.
[69] M. Skelton. (2016). *How to break apart a monolithic system safely without destroying your team, Available at: goo.gl/pqBVm2 [Last accessed: 4 November 2016].*
[70] *Self-Contained Systems: Assembling Software from Independent Systems, Available at: http://scs-architecture.org/ [Last accessed: 1 June 2017].*
[71] N. Dragoni *et al.*, "Microservices: Yesterday, Today, and Tomorrow," in *Present and Ulterior Software Engineering*, M. Mazzara and B. Meyer, Eds. Cham: Springer International Publishing, 2017, pp. 195-216.
[72] R. Chris. (2014). *Pattern: Monolithic Architecture, Available at: goo.gl/royZ7i [Last accessed: 4 November 2016].*
[73] G. Arun. (2015). *Microservices, Monoliths, and NoOps, Available at: goo.gl/zou2x3 [Last accessed: 8 November 2016].*
[74] S. Gibson. *Monoliths are Bad Design... and You Know It, Available at: goo.gl/xVEbSE [Last accessed: 4 March 2016].*
[75] M. E. Conway, "How do committees invent?," *Datamation,* vol. 14, no. 5, 1968.
[76] J. Humble. (2011). *Organize software delivery around outcomes, not roles: continuous delivery and cross-functional teams, Available at: goo.gl/MnFtJN [Last accessed: 10 August 2016].*
[77] P. Beijer and T. de Klerk, *IT Architecture - Essential Practice for IT Business Solutions*. Lulu.com, 2010.
[78] V. Gitlevich and E. Evans. *What is Domain-driven design? Available at: goo.gl/S3zMSR [Last accessed: 21 June 2016].*
[79] E. Evans, *Domain-driven design: tackling complexity in the heart of softwareT*. Addison-Wesley Professional, 2004.
[80] Alberto Brandolini. (2013). *Introducing Event Storming, Available at: goo.gl/GMzzDv [Last accessed: 8 July 2017].*
[81] W. Hasselbring and G. Steinacker, "Microservice Architectures for Scalability, Agility and Reliability in E-Commerce," in *2017 IEEE International Conference on Software Architecture Workshops (ICSAW)*, 2017, pp. 243-246.
[82] N. Ford. *Architecture is abstract until operationalized, Available at: goo.gl/HorpbH [Last accessed: 21 February 2016].*
[83] M. Fowler. (2015). *MicroservicePremium, Available at: goo.gl/3WVKsn [Last accessed: 31 October 2016].*
[84] I. Manotas *et al.*, "An empirical study of practitioners' perspectives on green software engineering," in *38th International Conference on Software Engineering*, Austin, Texas, 2016, pp. 237-248: ACM.
[85] B. Sokhan. *Domain Driven Design for Services Architecture, Available at: goo.gl/ftCLnR [Last accessed: 10 January 2016].*
[86] M. Erder and P. Pureur, *Continuous architecture: sustainable architecture in an agile and cloud-centric world*. Morgan Kaufmann, 2015.
[87] M. T. Rahman, L. P. Querel, P. C. Rigby, and B. Adams, "Feature Toggles: Practitioner Practices and a Case Study," in *2016 IEEE/ACM 13th Working Conference on Mining Software Repositories (MSR)*, 2016, pp. 201-211.
[88] T. d. Pauw. (2017). *Feature Branching is Evil, Available at: https://speakerdeck.com/tdpauw/xp2017-feature-branching-is-evil/ [Last accessed: 27 May 2017].*
[89] K. Gabhart. (2014). *Resilient IT Through DevOps, Available at: https://puppet.com/blog/resilient-it-through-devops/ [Last accessed: 1 July 2017].*
[90] K. Wnuk, "Involving Relevant Stakeholders into the Decision Process about Software Components," in *2017 IEEE International Conference on Software Architecture Workshops (ICSAW)*, 2017, pp. 129-132.
[91] N. Ernst, J. Klein, G. Mathew, and T. Menzies, "Using Stakeholder Preferences to Make Better Architecture Decisions," in *2017 IEEE International Conference on Software Architecture Workshops (ICSAW)*, 2017, pp. 133-136.
[92] S. Suneja *et al.*, "Safe Inspection of Live Virtual Machines," in *13th ACM SIGPLAN/SIGOPS International Conference on Virtual Execution Environments*, Xi'an, China, 2017, pp. 97-111, 3050766: ACM.
[93] L. Prewer. (2015). *Smoothing the continuous delivery path – a tale of two teams, Available at: goo.gl/1oqjsP [Last accessed: 2 October 2016].*
[94] D. Schauenberg. (2014). *Development, Deployment and Collaboration at Etsy, Available at: goo.gl/umGTM2 [Last accessed: 1 September 2017].*
[95] Y. Yaniv. (2014). *Closing the Gap Between Database Continuous Delivery and Code Continuous Delivery, Available at: goo.gl/mERZcV [Last accessed: 21 August 2016].*
[96] A. Brown. (2015). *What's the Best Team Structure for DevOps Success? Available at: goo.gl/3Z11og [Last accessed: 13 September 2017].*
[97] *What Team Structure is Right for DevOps to Flourish, Available at: goo.gl/KM6N3p [Last accessed: 24 September 2017].*
[98] M. Shahin, M. Zahedi, M. A. Babar, and L. Zhu, "Adopting Continuous Delivery and Deployment: Impacts on Team Structures, Collaboration and Responsibilities," in *21st International Conference on Evaluation and Assessment in Software Engineering*, Karlskrona, Sweden, 2017, pp. 384-393: ACM.
[99] J. Bosch, "Speed, Data, and Ecosystems: The Future of Software Engineering," *Software, IEEE,* vol. 33, no. 1, pp. 82-88, 2016.
[100] A. Wallgren. (2015). *Continuous Delivery of Microservices: Patterns and Processes, Available at: goo.gl/Yk6ddH [Last accessed: 10 February 2018].*
Preprint - accepted to be published in Empirical Software Engineering (2018)

31